\documentclass[12pt,a4paper]{article}
\usepackage[utf8]{inputenc}
\usepackage[english]{babel}
\usepackage{amsmath}
\usepackage{amsfonts}
\usepackage{amssymb}
\usepackage{graphicx}
\usepackage{jheppub}
\usepackage{cancel}
\usepackage{color}
\usepackage{slashed, epsf, latexsym}
\usepackage{epsfig}
\usepackage{graphics}

\newcommand{\U}{\Upsilon}

\usepackage{color}

\newcommand{\CG}{{\cal G}}
\newcommand{\CL}{{\cal L}}

\begin{document}
	\preprint{}
	\title{Stationary Solutions from the Large $D$ Membrane Paradigm}

	\author[a,b] {Mangesh Mandlik,}
	\author[c]{and Somyadip Thakur}

	\affiliation[a]{Department of Physics, Technion - Israel Institute of Technology, Haifa 3200003, Israel}
\affiliation[b]{Department of Mathematics, University of Haifa, Haifa 31905, Israel}
	\affiliation[c]{CAS Key Laboratory of Theoretical Physics, Institute of Theoretical Physics, Chinese Academy of Sciences, Beijing 100190, China
	}

	\emailAdd{ 	mangesh@physics.technion.ac.il}

	\emailAdd{somyadip@itp.ac.cn}
	
	\abstract{	It has recently been shown that the dynamics of black holes in large number of dimensions $D$ can be recast   as the  dynamics of a probe membrane propagating in the background  spacetime which solves Einstein equations without matter. The equations of motion of this membrane are simply the statement of conservation of the stress tensor and charge current defined on this membrane.
	 In this paper we obtain the effective equations of motion for stationary membranes in any empty background both in presence and absence of charge.  It turns out that the thermodynamic quantities associated with the stationary membranes that satisfy these effective equations also satisfy the first law of black hole thermodynamics. These stationary membrane equations have some interesting solutions such as  charged rotating black holes in flat and AdS backgrounds as well as  black ring solutions in large $D$.}
	
	\maketitle
	
\section{Introduction}

\hspace{1cm} The classical dynamics of $D$ dimensional spacetime is governed by Einstein's theory in which matter influences spacetime via the celebrated Einstein's equation
\begin{equation}
R_{MN} - \frac{R}2 G_{MN} + \frac{D-1}\lambda G_{MN} = 8\pi T_{MN}
\end{equation}
(in the natural units $c = G_N = e = 1$). The dynamical degrees of the spacetime are its metric $G_{MN}$. However even in the absence of matter ($T_{MN}=0$), it is one of the hardest equations to solve for a generic solution. This is because the curvature tensor $R_{MN}$ depends on the inverse of the metric tensor, which needs to be written in terms of metric components using the minors and the determinant, and ultimately we get $D(D+1)/2$ coupled differential equations of $D(D+1)/2$ functions of coordinates, which consist of rational functions of the metric components and their derivatives.\\

The diffeomorphism gauge symmetry of this equation can be exploited to choose a clever coordinate system which enables us to obtain exact solutions in highly symmetric static or stationary situations. This equation can also be linearized to get weak gravitational solutions such as gravitational waves. However these techniques can not be generalized to interesting dynamical situations like, say, collision and merger of two black holes. For these, one needs to rely on solving this equation numerically, and the nonlinearity is quite demanding on the computational power.\\

Recently the authors of \cite{Emparan:2013moa,Emparan:2013xia,Emparan:2013oza,Emparan:2014aba} made an observation that in the limit where $D \gg 1$, the degrees of freedom of the black hole spacetime separate into light degrees with length scale $r_0$ and heavy degrees with length scale $\frac{r_0}D$, where $r_0$ is a characteristic length scale associated with black hole horizon. They also showed in \cite{Emparan:2014cia,Emparan:2015rva} that the quasinormal modes about a static, spherically symmetric black hole in such a large number of dimensions show a similar separation of scales. So it should be possible to `integrate out' the heavy modes and get an effective theory of light modes which are way fewer in number.\footnote{See \cite{Emparan:2014jca,Emparan:2015hwa,Emparan:2016sjk,Tanabe:2016pjr,Tanabe:2016opw} for related discussion.}\\

In \cite{Bhattacharyya:2015dva} a systematic program of obtaining an effective theory of the light degrees of freedom of large $D$ black holes was initiated and named `a membrane paradigm at large $D$'. It was found out that $1/D$ becomes a new perturbative parameter in solving Einstein's equation for black hole solutions which are symmetric in all but a finite number of directions. If we start with a particular very simple ansatz for a black hole metric that solves the Einstein's equation without matter and cosmological constant at the leading order in $1/D$, and try solving for the perturbative corrections, it is observed that the black hole horizon acts like a codimension one `membrane' embedded in a flat $D$ dimensional background and characterized by its shape as embedded in the background and a `velocity field' defined on it. the dynamics of these shape and velocity is governed by the so called `membrane equations'. As the quasinormal modes about a static, spherically symmetric membrane match exactly with the light quasinormal modes obtained in \cite{Emparan:2014aba}, this membrane paradigm is indeed an effective theory of the light modes.\\

The perturbative expansion in $1/D$ indeed looks similar to the long wavelength expansion of Hydrodynamics, but there are many differences emerging from the nonlinearity of membrane equations at any given order. As a result, qualitative features of black hole like entropy production are visible even at the leading order. \cite{Dandekar:2016fvw} solve the next order in the perturbation and find the entropy production, indicating this membrane obeys the second law of thermodynamics like a black hole does. \cite{Bhattacharyya:2015fdk} addresses a membrane paradigm for charged asymptotically flat back holes, \cite{Bhattacharyya:2017hpj,Bads2} deal with uncharged black hole membranes in spacetimes with cosmological constant while \cite{Bhattacharyya:2018ads} works with the charged black hole membranes in spacetimes with cosmological constant. Also, \cite{Bhattacharyya:2016nhn} has shown that the membrane equations that dictate the dynamics of the charged membranes in a flat background are actually a statement of conservation of a stress tensor and a charge current associated with the membrane. \footnote{See \cite{Sadhu:2016ynd,Dandekar:2016jrp,Dandekar:2017aiv} for more work in this membrane paradigm.} \footnote{See \cite{Herzog:2016hob,Rozali:2016yhw,Chen:2016fuy,Chen:2017wpf,Chen:2017hwm,Rozali:2017bll,Chen:2017rxa,Raman:2017rfv,Herzog:2017qwp,Emparan:2018bmi,Chen:2018vbv,Andrade:2018zeb} for other work that uses large $D$ expansion.} \\

The equations that are obtained at any order in $1/D$ are still a system of nonlinear differential equations, but the advantage of this approach over solving the Einstein's equations with brute force is that membrane equations don't involve an inverse of some dynamical matrix, so the equations written in terms of the basic degrees of freedom of the membrane are `less nonlinear', and hence will be less computationally taxing when this formalism is developed for highly dynamical processes.\footnote{This comparison of the large $D$ membrane paradigm is with a naive numerical solution of Einstein's equations only. and not with the techniques used in the field of numerical relativity, for example at LIGO, which involve some very smart approximations that save a lot of computation.}\\

Though this membrane paradigm is potentially very useful for improving the efficiency of numerical black hole solutions, it would be very satisfying if it could give some analytic black hole solutions, at least perturbatively in $1/D$. An independent approach has led the authors of \cite{Suzuki:2015iha} to the conclusion that the stationary uncharged asymptotically flat or AdS black holes can be represented as membranes at the leading order in $1/D$ described by a very simple equation. So we can expect the leading order membrane equation presented in \cite{Bhattacharyya:2015fdk} or \cite{Bhattacharyya:2018ads} to reduce to the same equation when specialised to a stationary setting. The primary goal of this paper is to seek for the membrane equations for charged membrane in flat and AdS backgrounds specialised to general stationary configurations. These equations turn out to be very simple as expected, and reduce further to the equation in \cite{Bhattacharyya:2018ads} in the uncharged limit. These equations cannot be solved exactly in the simple examples we probe, namely a charged membrane rotating in a single plane, but can be solved perturbatively with ease. Also, these equations turn out to govern the thermodynamics of these membranes, which is quite expected because the equations we started from are the conservation of stress tensor and the charge current on the membrane.\\

This paper is organized as follows. In section \ref{secstat} we specialize the charged membrane equations for stationary configurations. In \ref{kill} we establish which class of membranes corresponds to stationary configurations, and in \ref{subsecstat} we find out the effective `stationary membrane equations' that describe such membranes. In section \ref{thermo} we describe the thermodynamics of these stationary membranes. The effective stress tensor and charge current on these membranes are proposed in \ref{EffcurrST}, thermodynamic data is extracted from them in \ref{flaw} and this data is shown to satisfy the first law of thermodynamics, defining the temperature and chemical potential associated with the membrane in the process.\\

Then in section \ref{rotmem}, we obtain the axially symmetric membrane solutions in flat and AdS backgrounds. We find exact solutions for uncharged membranes in both of these cases, while we obtain the charged counterparts perturbatively. In section \ref{rottherm} we show how the thermodynamics of \ref{thermo} is realized for the rotating membranes. The algebraic details relevant to this paper are given in the appendices in the end.

\section{Membrane equations for stationary configurations}\label{secstat}

\hspace{1cm} Before we begin, we give a quick overview of the notation used in this paper. Let's denote the coordinates of the background spacetime by $\{X^M\}$ and the coordinates on the membrane world volume by $\{y^{\mu}\}$. The upper case Latin indices are reserved for the quantities that transform under the spacetime coordinate transformations, while the lower case Greek indices are for the quantities that transform under the membrane coordinate transformations. The membrane covariant quantities are written in terms of spacetime covariant quantities using the pullback factors $e^M_{\mu} \equiv \frac{\partial X^M}{\partial y^{\mu}}$ as

\begin{equation}
T_{\mu_1\cdots\mu_n} = T_{M_1\cdots M_n}e^{M_1}_{\mu_1}\cdots e^{M_n}_{\mu_m},
\end{equation}
while the spacetime contravariant quantities are written in terms of membrane contravariant quantities as

\begin{equation}
T^{M_1\cdots M_n} = T^{\mu_1\cdots\mu_n}e^{M_1}_{\mu_1}\cdots e^{M_n}_{\mu_m}.
\end{equation} 

A large $D$ black hole membrane is characterized by:
\begin{itemize}
	\item Its shape as embedded in the background spacetime, i.e. the shape function $\rho(X) = 1$. The quantities that appear here which encode this shape function are a unit (outward) normal $~n_M \equiv N\partial_M\rho~$ where $~N^{-2} = \partial_M\rho\partial^M\rho~$, extrinsic curvature tensor $K_{MN}$ (with trace $K$) and its derivatives. The lowering and raising are done by the background metric $G_{MN}$ and its inverse $G^{MN}$ respectively.
	\item A unit normalized velocity field $u^M$, $u^Mu_M = -1$ that lies in the membrane, i.e. $u^M n_M = 0$.
	\item A scalar charge field $Q(y)$.
\end{itemize}

The vector membrane equation has components entirely in the membrane. The constituents of both the vector and the scalar membrane equations such as the extrinsic curvature, the velocity and charge field also lie entirely in the membrane. Therefore in this paper we will work with quantities as a function of $\{y^\mu\}$ and they will transform under membrane coordinate transformations, i.e. they will have Greek lower case indices.
Horizon Tunneling Revisited: The Case of Higher Dimensional Black
Holes
\begin{equation}
\begin{split}
g_{\mu\nu} &= G_{MN}e^M_{\mu}e^N_{\nu} ~~~~~~~~\text{(Membrane metric)},\\
K_{\mu\nu} &= K_{MN}e^M_{\mu}e^N_{\nu},\\
u_\mu &= u_Me^M_{\mu},\\
p_{\mu\nu} &= g_{\mu\nu} + u_\mu u_\nu~~~~~~~~\text{(Projector orthogonal to velocity)}.
\end{split}
\end{equation}

Also, whenever the covariant derivatives and Riemann and Ricci curvature tensors are seen wearing hats, they should be assumed to be taken w.r.t. the membrane metric. Their unhatted counterparts are defined w.r.t. the background metric.

\subsection{Large $D$ membrane equations}\label{memeq}

The charged membrane equations in presence of a cosmological constant are obtained in \cite{Bhattacharyya:2018ads}, of which the vector equation is:
\begin{equation}\label{VeqAdS}
{\cal E}_\mu \equiv \left(\frac{\hat{\nabla}^2 u_\nu}{K}-(1-Q^2)\nabla_\nu \ln K + u\cdot K_\nu -(1+Q^2)u\cdot\hat{\nabla}u_\nu\right)p^\nu_\mu = 0,
\end{equation}

while the scalar equation is:
\begin{equation}\label{SeqAdS}
{\cal E} \equiv \frac{\hat{\nabla}^2 Q}{QK}-u\cdot\nabla \ln Q - u\cdot\nabla\ln K +u\cdot K \cdot u+\frac{u\cdot R \cdot u}{K} = 0.
\end{equation}

In the flat limit ($R_{MN} = 0$), the equations reduce to those in \cite{Bhattacharyya:2015fdk}. In uncharged limit ($Q = 0$) the vector equation reduces to the membrane equation in \cite{Bhattacharyya:2017hpj}, while the scalar equation vanishes identically. This and the next section of this paper deal with stationary membranes in empty backgrounds with arbitrary cosmological constant (with magnitude ${\cal O}(D)$ or smaller) including zero, and any arbitrary charge (with $|Q| \sim {\cal O}(1)$ or smaller) including, again, zero. However in the section after that, \ref{rotmem}, where we specialize to the particular examples of rotating membrane solutions, we work with backgrounds of nonpositive cosmological constant, i.e. flat and AdS backgrounds, and we treat the uncharged and charged cases separately in each of them.

\subsection{Time Like Killing Vectors and Stationary Membrane velocity configuration}\label{kill}

 \hspace{1cm} As mentioned in the introduction, this paper deals with stationary solutions of the membrane equations \eqref{VeqAdS} and \eqref{SeqAdS}. But what defines the stationary solutions?\\
 
The membrane equations that dictate the dynamics of the large $D$ black hole membrane are actually the equations of conservation of charge current and stress tensor defined on the membrane, as shown in \cite{Bhattacharyya:2016nhn} \footnote{The result in \cite{Bhattacharyya:2016nhn} was obtained for a charged membrane in flat space, but the analysis can be extended to a charged membrane in AdS (or dS)}. This current and stress tensor can be seen as belonging to a fluid living on the membrane world volume \footnote{This statement is not quite accurate. According to \cite{Dandekar:2017aiv} the membrane stress tensor is a fluid stress tensor plus the Brown York stress tensor of the membrane due to its embedding in the background. However the Brown York part is automatically conserved, plus it doesn't contribute to the dissipative part of the stress tensor which is our focus in this argument.}. So if we look at the stress tensor of this `fluid'

\begin{equation}\label{effST}
8\pi T_{\mu\nu}^{(eff)} =\left( \frac{1+Q^2}{2} \right) K u_\mu u_\nu + \left( \frac{1-Q^2}{2}  \right)K_{\mu\nu}- \Sigma_{\mu\nu}
-Q\left( u_\mu{\cal V}_\nu + u_\nu{\cal V}_\mu\right),
\end{equation}
where
\begin{equation}\label{visc}
\begin{split}
\Sigma_{\mu\nu} &= \tilde{\Sigma}_{\mu\nu} - \tilde{\Sigma}^\alpha_\alpha p_{\mu\nu},\\
\tilde{\Sigma}_{\mu\nu} &= p^{\alpha}_{\mu}\left(\frac{\hat\nabla_\alpha u_\beta + \hat \nabla_\beta u_\alpha}{2} \right)p^{\beta}_{\nu},
\end{split}
\end{equation}
and
\begin{equation}\label{calv}
{\cal V}_\mu =\hat\nabla_\mu Q +Qu\cdot\hat{\nabla}u_{\mu},
\end{equation}
we can make out the shear viscosity term $\Sigma_{\mu\nu}$. \footnote{The bulk viscosity is subleading at the relevant order in $\frac{1}D$.} This suggests that the dynamics of the membrane is dissipative, and a general dynamical membrane will eventually settle down to a stationary configuration, and it will be characterized by disappearance of the dissipation term. Thus for the stationary configuration, $\Sigma_{\mu\nu} = 0$.\\

Now consider the case in which the membrane has a time like Killing vector field $k^{\mu}$. Let us define a unit velocity field $u^{\mu}$ proportional to this time like Killing vector field as $$u^{\mu}=\gamma k^{\mu},$$
where $$\gamma^{-2} = -k_{\mu}k^{\mu}.$$
If we identify this velocity field with the velocity field that defines the membrane, we have then

\begin{equation}\label{Sigstat}
\begin{split}
\tilde{\Sigma}_{\mu\nu} &= p_{\mu}^{\alpha}p_{\nu}^{\beta}\left(\hat{\nabla}_{\alpha}u_{\beta}+\hat{\nabla}_{\beta}u_{\alpha}\right),\\
&= p_{\mu}^{\alpha}p_{\nu}^{\beta}\left(\nabla_{\alpha}(\gamma k_{\beta})+\nabla_{\beta}(\gamma k_{\alpha})\right),\\
&= p_{\mu}^{\alpha}p_{\nu}^{\beta}\left(k_{\beta}\nabla_{\alpha}\gamma +k_{\alpha}\nabla_{\beta}\gamma \right),\\
&= 0.
\end{split}
\end{equation}

Where going from second line to the third we have used the Killing equation and then we have used the fact that $p_{\mu\nu}$ is a projector orthogonal to $u^{\mu}$ ,and hence orthogonal to $k^{\mu}$. \eqref{Sigstat} means $\Sigma_{\mu\nu} = 0$.\\

Let's appreciate the significance of this result for a moment. First we choose a timelike Killing vector $k(X)$ of the background, Since this generates an isometry of the empty background, it is also the symmetry of this background. Then let's construct a membrane shape in this background that respects this symmetry. Now look at the restriction of $k(y)$ on the membrane. Since the membrane breaks any symmetry in the direction orthogonal to it, this restriction shouldn't have a component normal to the membrane. $k(y)$ turns out to be the Killing vector of the metric induced on the membrane as well because
$$\hat{\nabla}_\mu k_\nu+\hat{\nabla}_\nu k_\mu = e^M_\mu\left(\nabla_Mu_N+\nabla_Nu_M\right)e^N_\nu = 0.$$
Then we unit normalize $k(y)$ to get $u(y)$, and choose $u(y)$ to be the velocity field of the membrane.\footnote{Since $k$ doesn't have a component normal to the membrane, it doesn't matter it the normalization is done w.r.t. the background metric or the induced metric on the membrane.} \eqref{Sigstat} ensures that a membrane constructed in this way is stationary.\footnote{The association of a Killing vector with a stationary configuration has been already derived in \cite{Caldarelli:2008mv} in the context of a fluid (plasma), but the derivation here is much simpler and more relevant to the membrane picture.}\\

Also, physically it makes sense that the charge field also should have the symmetry generated by $k$, making $k$ the generator of a symmetry of the membrane. However, this fixes neither the embedding of the membrane in the background spacetime nor the charge field as a function of the non-symmetry directions entirely, for that we have to solve the membrane equations.\\

\subsection{Stationary Membrane equations}\label{subsecstat}

\hspace{1cm} The previous subsection enables us to fix the velocity field on the membrane, so now there are only two unknown functions left, namely the shape function and the charge density function. On the other hand the number of independent equations to be solved is still $D-2$. For the system to be solvable, these equations must boil down to just two independent equations, and to respect diffeomorphism invariance, they must be scalar equations.\\

Using \eqref{uKupuRu} for stationary configurations, the scalar equation \eqref{SeqAdS} reduces to

\begin{equation}\label{Sstat}
\tilde{\nabla}^2\ln\frac{Q}{\gamma} = Ku\cdot\hat{\nabla}\ln(QK).
\end{equation}

Using \eqref{tnabsqupuK} and \eqref{gamp2}, and rearranging terms we convert the vector equation \eqref{VeqAdS} into

\begin{equation}\label{Vstat}
(1-Q^2)\tilde{\nabla}_\mu\ln K - (1+Q^2)\tilde{\nabla}_\mu\ln\gamma = ((1-Q^2)u\cdot\hat{\nabla}\ln K - (1+Q^2)u\cdot\hat{\nabla}\ln\gamma))u_\mu.
\end{equation}

Since the membrane is symmetric along $u$, the RHS of both \eqref{Sstat} and \eqref{Vstat} vanish. 
From \eqref{Sstat} we get, assuming reasonable boundary conditions on the compact and extended directions of the membrane
\begin{equation}\label{Qgam}
Q = \alpha\gamma,
\end{equation}
where $\alpha$ is a constant. Substituting \eqref{Qgam} in  \eqref{Vstat} while setting the RHS of \eqref{Vstat} to zero, and an easy manipulation \eqref{vmanip} gives
\begin{equation}
\tilde{\nabla}_\mu\ln\left(K\left(\frac{1}{\gamma}-\alpha^2\gamma\right)\right) = 0.
\end{equation}
Which can be integrated over the membrane, and ultimately we get
\begin{equation}\label{Statsol}
\begin{split}
Q &= 2\sqrt{2\pi}\mu\gamma,\\
K &= \frac{4\pi T\gamma}{1-\alpha^2\gamma^2} = \frac{4\pi T\gamma}{1-Q^2}~~,
\end{split}
\end{equation}
$T$ and $\mu$ are constant all over the membrane. The reason behind putting the seemingly arbitrary numerical factors in these equations will be clear when we look at the thermodynamic of the stationary membranes.\\

In the absence of charge, the stationary membrane equations \eqref{Statsol} reduce to

\begin{equation}\label{Statsolunch}
K = 4\pi T\gamma
\end{equation}

\section{Thermodynamics of stationary membranes}\label{thermo}

\subsection{Effective stress tensor and charged current}\label{EffcurrST}

\hspace{1cm} The effective stress tensor and charge current belong to a family of stress tensors and charge currents the conservation of which gives the membrane equations at the leading order. We define those by simply covariantizing the ones defined in equations (6.19) and (6.34) in \cite{Bhattacharyya:2016nhn}. The actual stress tensor and charge current need to be found out by the procedure detailed in \cite{Bhattacharyya:2016nhn}. However for our purpose we only need the leading order stress tensor and charge current, and since the actual and effective quantities differ by subleading terms which don't contribute even to the membrane equations, the leading parts can equally be read off from the effective quantities. The colour coding used in the equations in this section is explained in the appendix \ref{idens2}

\subsubsection{The effective charge current and its conservation}\label{eccc}

\hspace{1cm} We propose, as discussed above, the effective current on the membrane to be

\begin{equation}\label{effcurr}
2\sqrt{2\pi} J_\mu^{(eff)}
=~Q {K} u_\mu -{\cal V}_\mu,
\end{equation}
where ${\cal V}$ is defined in \eqref{calv}.

Therefore taking its divergence gives
\begin{equation}\label{effcurrcons}
\begin{split}
2\sqrt{2\pi} \hat{\nabla}^{\mu}J_\mu^{(eff)} =& u\cdot\nabla(KQ) \textcolor{blue}{+KQ\hat{\nabla}\cdot u}- \hat{\nabla}\cdot{\cal V}\\
=& ~u\cdot\nabla(KQ) - \hat{\nabla}^2Q - u\cdot R\cdot u - K~u\cdot K\cdot u + {\cal O}(1)\\
=& -KQ\left(\frac{\hat{\nabla}^2 Q}{QK}-u\cdot\nabla \ln Q - u\cdot\nabla\ln K +u\cdot K \cdot u+\frac{u\cdot R \cdot u}{K}\right)\\ &+ {\cal O}(1).
\end{split}
\end{equation}
Here we have used \eqref{Vid}. If we demand this current to be conserved, i,e, its divergence to vanish, \eqref{effcurrcons} implies the scalar membrane equation \eqref{SeqAdS}.

\subsubsection{The effective stress tensor and its conservation}\label{estc}

When we take the divergence of the effective stress tensor \eqref{effST} on the membrane, we get
\begin{equation}\label{effSTcons}
\begin{split}
8\pi \hat{\nabla}^\mu T_{\mu\nu}^{(eff)} =& \left( \frac{1+Q^2}{2} \right) Ku\cdot\hat{\nabla}u_{\nu} + \left( \frac{1+Q^2}{2} \right) (u\cdot\nabla K) u_{\nu}+KQ(u\cdot\nabla K) u_{\nu}\\
&+ \textcolor{blue}{\left( \frac{1+Q^2}{2} \right) K(\hat{\nabla}\cdot u)u_{\nu}}+\left( \frac{1-Q^2}{2}  \right)\hat{\nabla}^\mu K_{\mu\nu} \textcolor{blue}{+Q(\nabla^\mu Q) K_{\mu\nu}}\\
&- \hat{\nabla}^{\mu}\Sigma_{\mu\nu} - Q(\hat{\nabla}\cdot{\cal V}) u_{\nu} \textcolor{blue}{-Q(\hat{\nabla}\cdot u){\cal V}_{\nu}-Qu\cdot\hat{\nabla}{\cal V}_{\nu}-Q{\cal V}\cdot\hat{\nabla}u_\nu}\\
&-\textcolor{blue}{(u\cdot\nabla Q){\cal V}_\nu - ({\cal V}\cdot\hat\nabla Q)u_\nu}\\
=& -\frac{K}2\left(\frac{\tilde{\nabla}^2 u_\mu}{K}-(1-Q^2)\nabla_\mu \ln K + u\cdot K_\mu -(1+Q^2)u\cdot\tilde{\nabla}u_\mu\right)\mathcal{P}^\mu_\nu\\
&- KQ^2\left(\frac{\hat{\nabla}^2 Q}{QK}-u\cdot\nabla \ln Q - u\cdot\nabla\ln K +u\cdot K \cdot u+\frac{u\cdot R \cdot u}{K}\right)u_{\nu}\\ &+{\cal O}(1).
\end{split}
\end{equation}
Here we have used \eqref{Bid3} and \eqref{Vid}. If we demand this stress tensor to be conserved, i,e, its divergence to vanish, the components of \eqref{effSTcons} orthogonal to $u$ imply the vector membrane equation \eqref{VeqAdS}, whereas the component of \eqref{effSTcons} along $u$ implies the scalar membrane equation \eqref{VeqAdS}.

\subsubsection{The uncharged limit}\label{unlim}\hspace{1cm}
Setting $Q = 0$ in \eqref{effcurr} makes it identically vanish, while \eqref{effST} gives in this special case
\begin{equation}
16\pi T_{\mu\nu}^{Q=0} = K u_\mu u_\nu + K_{\mu\nu}- 2\Sigma_{\mu\nu}
+ {\cal O}\left(\frac{1}{D}\right),
\end{equation}
which matches with the stress tensor of the uncharged membrane proposed in \cite{Dandekar:2017aiv}.\\

In short, the uncharged limit and the flat limit of \eqref{effcurr} and \eqref{effST} agree with the known result, while their conservation gives the membrane equations. We take it as a sufficient evidence for these to be the AdS membrane effective charge current and stress tensor, and proceed to compute thermodynamic quantities using them. We will show that   these quantities satisfy the first law of thermodynamics and  thus further bolstering this claim.

\subsection{Thermodynamic quantities from membrane data}\label{unch1stlaw}

\hspace{1cm} In this section we derive the thermodynamic quantities associated with a stationary membrane from the membrane data, i.e, the stress tensor,  charge current and the entropy current. We show that these thermodynamic quantities obey the first law of black hole thermodynamics.\\

We can define a conserved current from the leading order stress tensor as \footnote{In \eqref{concur} we have defined the conserved current $J_{(E)}^{\mu} $ with a negative sign so that the energy density  which is the $J_{(E)}^0$ component of the conserved current, positive. }
\begin{equation}
\label{concur}
J_{(E)}^\mu = -T^{\mu\nu}k_\nu = (1+Q^2)\frac{K}{16\pi\gamma}u^\mu.
\end{equation}
Now let's consider a spacelike slice of the membrane world volume with normal along $dt$. The conserved `total energy' on this slice is given by

\begin{equation}\label{memtoten}
\tilde{E} = \int d^{D-2}x \sqrt{-G} J_{(E)}^0 = \int d^{D-2}x \sqrt{-G} (1+Q^2)\frac{K}{16\pi},
\end{equation}
since $u^0 = \gamma$. Note that this `total energy' also includes the contributions from angular and linear momenta if $k$ has contributions from the rotation and spatial translation Killing vectors respectively. 

A conserved charge is defined on such a slice from the leading order charge current $J^\mu = \frac{KQ}{2\sqrt{2\pi}}u^\mu$

\begin{equation}\label{memcharge}
q = \int d^{D-2}x \sqrt{-G} J^0 = \int d^{D-2}x \sqrt{-G}\frac{KQ\gamma}{2\sqrt{2\pi}}.
\end{equation}

And finally the entropy is simply

\begin{equation}\label{mementropy}
S = \int d^{D-2}x \sqrt{-G}\frac{u^0}{4} = \int d^{D-2}x \sqrt{-G}\frac{\gamma}{4}.
\end{equation}

All the integrations are taken over the aforementioned slice of the membrane worldvolume, $\{x\}$ are the $D-2$ coordinates on that slice. 

\subsection{First law of thermodynamics}\label{flaw}
\hspace{1cm} Now let's assume that the relative variations in $Q$, $\gamma$ and $K$ are of the same order, i.e. $$\frac{\delta Q}{Q} \sim \frac{\delta K}{K} \sim \frac{\delta \gamma}{\gamma}.$$
Let $l$ be a typical length scale of the membrane. Note that $K \sim \frac{D}{l}$ while $\sqrt{-G} \sim {l^{D-2}}$. Thus, $\frac{\delta\sqrt{-G}}{\sqrt{-G}} \sim -D\frac{\delta K}{K}$. Thus when the variations of thermodynamic quantities \eqref{memtoten}, \eqref{memcharge} and \eqref{mementropy} are taken, the variation of the volume factor dominates over the variation of the quantities that are getting integrated.

So
\begin{equation}\label{vartherm}
\begin{split}
\delta S &= \int_{\rho=1} d^{D-2}x~ \delta\sqrt{-G}~\frac{\gamma}{4},\\
\delta \tilde{E} &= \int_{\rho=1} d^{D-2}x~ \delta\sqrt{-G}~(1+Q^2)\frac{K}{16\pi},\\
\delta q &= \int_{\rho=1} d^{D-2}x~ \delta\sqrt{-G}~\frac{KQ\gamma}{2\sqrt{2\pi}}.
\end{split}
\end{equation}

Now let's simplify the combination $T\delta S + \mu\delta q$, where $T$ and $\mu$ are defined in \eqref{Statsol}.
\begin{equation}\label{thermoproof}
\begin{split}
T\delta S +\mu\delta q &= \int_{\rho=1} d^{D-2}x~ \delta\sqrt{-G}\left(T\frac{\gamma}{4} + \mu\frac{KQ\gamma}{2\sqrt{2\pi}}\right),\\
&= \int_{\rho=1} d^{D-2}x~ \delta\sqrt{-G}\left(\frac{(1-Q^2)K}{4\pi\gamma}\frac{\gamma}{4} + \frac{Q}{2\sqrt{2\pi}\gamma}\frac{KQ\gamma}{2\sqrt{2\pi}}\right),\\
&= \int_{\rho=1} d^{D-2}x~ \delta\sqrt{-G}~(1+Q^2)\frac{K}{16\pi},\\
&= \delta \tilde{E}.
\end{split}
\end{equation}

Thus if we identify $T$ to be the membrane temperature and $\mu$ to be the membrane chemical potential, \eqref{thermoproof} is the statement that a stationary membrane satisfies the first law of thermodynamics at the leading order in large $D$.

\section{Stationary rotating membrane solutions}\label{rotmem}\hspace{1cm}
In section \ref{secstat}  we have derived the stationary  membrane equations which have solutions  dual to black holes in large D for the flat space, $AdS$-spaces both in presence and absence of charge. In this section we would specialize in solving the membrane equations for some axially-symmetric configurations. 
\subsection{The effective membrane equation for Stationary `axially-symmetric' configurations  in flat space}\hspace{1cm}
In this subsection we will find axially-symmetric solutions to \eqref{Statsol} and \eqref{Statsolunch} in the ambient flat space both in presence and absence of charge which will be dual to asymptotically flat rotating black holes in presence and absence of charge respectively.\\
A time like killing vector field of \eqref{flsplit}  which generates  the axially-symmetry  of the flat background and hence the final membrane configuration is given by
\begin{equation}
k=\frac{\partial}{\partial t}+\omega^{i}\frac{\partial}{\partial\theta_i}.
\end{equation}
Hence, the unit time-like velocity field of the membrane that we consider is given by
\begin{equation}\label{gamma}
u^{\mu}=\gamma k^{\mu},\quad \text{where} \quad \gamma=\frac{1}{\sqrt{\left(1-\sum_{i}r_i^2 \omega_i^2\right)}}.
\end{equation}
Now, the shape function of the surface that we consider should be such that its normal is orthogonal to the velocity field $u$. Also, we consider situation which preserves the $SO(D-2p-2)$ isometry of the metric \eqref{flsplit}. The surface equation that we consider is given by
\begin{equation}\label{shapeansatz2}
s^2=2g(\{r_i\}).
\end{equation}
The unit normal to the above surface is given by
\begin{equation}
n=2N\left(s ds-\frac{\partial g}{\partial r_i}\right), \quad \text{where,} \quad N=\frac{1}{2\sqrt{s^2+\sum_i\left(\frac{\partial g}{\partial r_i}\right)^2}}.
\end{equation}
The determinant of the metric \eqref{flsplit} is given by
\begin{equation}\label{metdetflt}
\sqrt{-G}=s^{D-2p-2}\sqrt{\Omega_{D-2p-2}}\prod_ir_i.
\end{equation}
Hence we can define the  trace of the extrinsic curvature of the surface \eqref{shapeansatz2} is given by
\begin{equation}
K=\nabla_{\mu}n^{\mu}=\frac{1}{\sqrt{-G}}\partial_{\mu}\left(\sqrt{-G}n^\mu\right).
\end{equation}
From \eqref{metdetflt} it can be easily seen that the leading contribution to the $K$ comes from  $s$ derivative acting  on the determinant of the metric and hence the trace of the extrinsic curvature is given by 
\begin{equation}\begin{split}\label{kappa}
K&=\frac{1}{\sqrt{-G}}n^s\partial_s(\sqrt{-G}),\\
K&=D\frac{n^s}{s}=D\frac{1}{\sqrt{s^2+\sum_i\left(\frac{\partial g}{\partial r_i}\right)^2}}.
\end{split}\end{equation}

\subsection{Rotating uncharged membrane solutions in flat background }
The equation of the membrane in a stationary configuration with no charge($Q=0$)  is given by\eqref{Statsolunch}
\begin{equation}\label{effleqn}
K=\frac{ \gamma}{\beta},
\end{equation}
where we have defined $\beta=\frac{D}{4\pi T}$.\\
Using the expressions of $K$ and $\gamma $ from (\eqref{gamma},\eqref{kappa}) we get the effective equation for the surface defined on the $s^2=2g$ as
\begin{equation}
\label{shapeflt}
2g+\sum_i(\partial_ig)^2
=\beta^2\left(1-\sum_ir_i^2\omega_i^2\right).
\end{equation}
Let us redefine $g, r_i$ and $\omega_i$ such that we can rewrite \eqref{shapeflt} in an intrinsic scale independent form.
\begin{equation}
\label{redsur}
g=\beta^2 \CG,\qquad r_i=\beta x_i \qquad \text{and} \qquad\omega_i=\frac{\U_i}{\beta}.
\end{equation}
In terms of these new variables we can write the equation of the membrane in the stationary configuration as 
\begin{equation}
\label{shapefltnw}
2\CG+\sum_i(\partial_i\CG)^2
=\left(1-\sum_ix_i^2\U_i^2\right),
\end{equation} where the derivative is taken w.r.t, $x_i$.\\
Let us now analyze the possible solutions of \eqref{shapefltnw}. One immediate observation is that if  $\CG$ is  at most quadratic in $x_i$ then so is every term in the equation shape and it forms a closed set of linear algebraic equations.
Moreover if the quadratic form in $\CG$ is diagonal in $i$ space (i.e; there are no $x_i x_j$ 
terms) then the same is true of every term in \eqref{shapefltnw}. So it is natural
to seek solutions to this equation of the form 
$$ \CG(x_i)= \sum_i\left(\frac{A_i}{2}  - \frac{C_i }{2}(x_i-B_i)^2\right).$$ 
For a non-trivial solution to \eqref{shapefltnw}  we require that  $C_i\neq0$ .\\
Equating the coefficients of the constant, linear and quadratic terms 
respectively in \eqref{shapefltnw} yields following set of linear algebraic  equations 
\begin{equation}\label{quadsolflt} \begin{split}
&\sum_i\left(A_i+B_i^2C_i(C_i-1)\right) = 1,\\
&B_iC_i(C_i-1)=0,\\
&C_i(C_i-1)+  \U_i^2=0.\\
\end{split}
\end{equation}
Let us now analyze the solutions to the set of equations in \eqref{quadsolflt}. We have two different sets of possible solutions which we list below,\\
\begin{itemize}
	\item For non-zero rotation $(\U_i\neq 0)$ : \\Since $C_i\neq0$, the consistent  set of solutions is
	\begin{equation}\begin{split}
	\sum_iA_i=1, \qquad B_i=0, \qquad \text{and}\qquad C_{i\pm}= \frac{1}{2} \left(1\pm \sqrt{1-4 \U_i ^2}\right).
	\end{split}
	\end{equation}
	\item For static solutions $(\U_i= 0)$ : \\Since $C_i\neq0$,  the possible set of solutions is \\
	\begin{equation}\begin{split}\label{sit2}
	\sum_iA_i=1, \qquad B_i=\text{free parameter},\qquad C_i=1.
	\end{split}
	\end{equation}
\end{itemize}
Let us now analyze each of these cases in some more detail.\\\\
{ \bfseries{\underline{ Rotating solutions $\U_i\neq0$}}} \\
In this case $A_i=1$, $B_i=0$ and 
\begin{equation}\label{csol}
C_{i\pm}= \frac{1}{2} \left(1\pm \sqrt{1-4   \U_i ^2}\right).
\end{equation}
The spacetime membrane is described by the equation
\begin{equation}\label{actsol}
s^2+ r_i^2 \frac{1 \pm \sqrt{1- 4  \U_i ^2}}{2} = \beta^2.
\end{equation}
This is an ellipsoidal membrane which is fatter in the $i$ directions than 
all other directions (this excess bulge is a consequence of the 
centrifugal force). These configurations are dual to Myers Perry type 
black hole solutions.
The above solution for the shape function is real and have a consistent solution for 
\begin{equation}\label{Upsrange}
-1/2< \U_i <1/2.
\end{equation}

\begin{figure}\begin{center}
	\includegraphics[width=120mm]{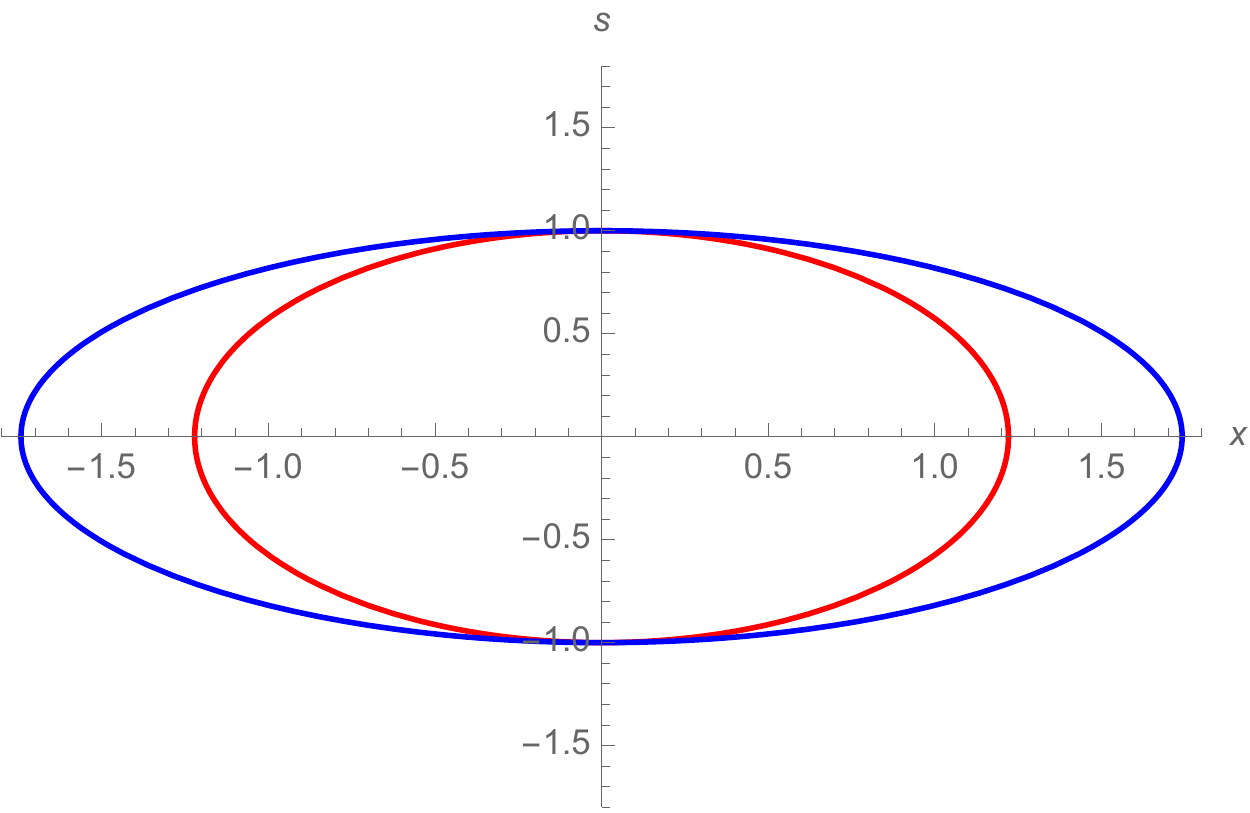}
	\caption{Uncharged Membrane surface  with rotation along one direction for   $\U=0.47$ for the $+$ and $-$ branch of \eqref{actsol}  }
	\label{fig:flat1}\end{center}
\end{figure}

Figure \ref{fig:flat1} shows the surface for turning on rotation along one plane . The blue curve corresponds to the $-$ branch of \eqref{actsol} and the red curve corresponds to the $+$ branch of \eqref{actsol}.

\newpage

\underline{{\bf{ Static Solutions$(\U_i=0)$}}} \\

In the static configuration , i.e; $\U_i=0$, we have a special branch of solution where 
$C_i=1$ and  $B_i$ is a free parameter. The membrane shape is given by 
the equation 
\begin{equation}\label{surfmanif}
s^2+ \sum_i\left(r_i-\beta B_i\right)^2= \beta^2.
\end{equation}
This does ${\it not}$ represent a spherical black hole with a shifted 
origin because the shift is in polar coordinates in  a two plane. 
The surface above is non analytic at $r_i=0$ (and therefore unacceptable) 
for $B_i^2< A_i$. \footnote{Where $A_i$ satisfies the constraint $\sum_iA_i=1$\eqref{sit2}.} On the other hand if $B_i^2> A_i$ then the point $r_i=0$ does not
lie on the manifold \eqref{surfmanif}. 
In fact \eqref{surfmanif} represents
a smooth black ring solution with horizon topology $(S^1)^n \times S^{D-2n-2}$ where we have non-zero black ring parameter $B_i$ turned on along $n$ two planes and $n<\frac{D-2}{2}$. 
The surface may be thought of as $S^{(D-2n-2)}$ fibered over direct product of $n$ annuli \footnote{See  \cite{Armas:2014bia} for perturbative uncharged black rings in flat space for any value of D. } . The inner radius of the annulus is $r_{i-}=\beta(B_i-\sqrt{A_i})$ and the outer radius is 
$r_{i+}=\beta(B_i+\sqrt{ A_i})$. The $S^{(D-2n-2)}$ shrinks to zero at both the inner and outer radius of the all the $n$ annuli $(r_{1\pm}, r_{2\pm},r_{3\pm}\cdots r_{n\pm})$.  \\

Notice that non rotating black rings exist only when their outer radius exceeds
$2\beta\sqrt{ A_i}$. At fixed $\beta\sqrt{A_i}$ we thus have a single solution with outer radius $\beta\sqrt{ A_i}$, 
and then solutions (ring solutions) for all values of the outer radius 
greater than $2\beta\sqrt{ A_i}$. \\

When $B_i\neq 0$ it describes a black ring solution as described above. However 
since $B_i$ is a free parameter we can have a possibility where  it can be set to zero. When  $B_i=0$  for all $i's$ the solution is spherical in all directions, and so represents the Schwarschild black hole. 
The spacetime membrane is described by the equation
\begin{equation}\label{actsschw}
s^2+ \sum_{i=0}^{p}r_i^2  = \beta^2.
\end{equation}
In the static configuration we can also have an interesting solution where we 
can set $B_i=0$ in  $(p-n)$ planes and in the rest $n$ planes it can be fixed non-zero constant parameter. 
The membrane shape is then given by  
\begin{equation}\label{bhring}
s^2+ \sum_{i=0}^{n}\left(r_i-\beta B_i\right)^2+\sum_{j=0}^{p-n}r_j^2=\beta^2. 
\end{equation}
For the flat space the stationary configurations  we can have solutions to the shape functions such that rotations are turned on in some planes and zero rotations in the rest of the  planes. 
So the most general solution where rotations are turned on $k$ planes and zero rotations in $(p-k)$ planes is given by 
\begin{equation}\label{genflt}
s^2+ \sum_{i=0}^{k}r_i^2 \frac{1 \pm \sqrt{1- 4  \U_i ^2}}{2}+\sum_{m=0}^{n}\left(r_{m}-\beta B_m\right)^2+\sum_{j=0}^{p-n-k}r_j^2 = \beta^2.
\end{equation}
where we have rotating black hole solution in $k$ planes, black ring solutions in $n$ and   non-rotating  black hole in $p-n-k$ directions.
\subsection{Rotating charged membrane solution in  flat background}
The most general solutions of the membrane in presence of charge in a stationary configurations is given by 
\begin{equation}\label{chflat}
\begin{split}
Q&=\alpha\gamma,\\
K&= \frac{\gamma}{\beta(1-Q^2)},
\end{split}
\end{equation}
where we have used the definition 
$$\alpha=2\sqrt{2\pi} \mu$$ and $$\beta=\frac{D}{4\pi T}.$$

The extrinsic curvature of the membrane as embedded in flat space is given by \eqref{kappa} and \eqref{gamma} and the  effective equation for the surface is given by 
\begin{equation}
\begin{split}
\label{efchflt}
& \left(2g+\sum_i(\partial_ig)^2\right)\left(1-\sum_ir_i^2\omega_i^2\right)\\
&=\beta^2\left(1-\sum_ir_i^2\omega_i^2-\alpha^2\right)^2.
\end{split}
\end{equation}
With the redefinition \eqref{redsur} we can rewrite \eqref{efchflt} in an intrinsically scale independent form as  
\begin{equation}
\begin{split}
\label{efchfltnw}
& \left(2\CG+\sum_i(\partial_i\CG)^2\right)\left(1-\sum_ix_i^2\U_i^2\right)\\
&=\left(1-\sum_ix_i^2\U_i^2-\alpha^2\right)^2.
\end{split}
\end{equation}
The static charged black hole and black ring solutions can be easily obtained by replacing  $\beta$ with  $\beta(1-\alpha^2)$ in the static uncharged case.\footnote{See  \cite{Armas:2014rva,Armas:2015qsv} for perturbative calculations of charged black rings in Einstein-Maxwell-Dilaton theory in flat space  AdS spacetime for arbitrary  D. }\\
The effective equations for the membrane in the stationary configuration is presence of charge is  hard to solve non-perturbatively. However we can solve the equation perturbatively in two different scheme
\begin{itemize}
	\item Perturbatively adding charge to neutral rotating membrane.
	\item Perturbatively adding rotation to static charged membrane.
\end{itemize}
{\bf{Perturbatively adding charge to neutral rotating membrane}}\\\\
We will solve the equation \eqref{efchfltnw}
for the simplest case where only one rotation is turned on. Let $(r,\theta)$ be the polar co-ordinates in this plane and $x=\frac{r}{\beta}$.  We will  do a perturbative expansion in the charge about a neutral  membrane rotating in one plane. Let us assume that the  expansion of the shape function in the charge parameter is given by 
\begin{equation}
\CG=   \CG_0 + \alpha^2   \CG_1 + \alpha^4   \CG_2 + \alpha^6   \CG_3 + \cdots
\end{equation}

Substituting in \eqref{efchfltnw} and separating out powers of $\alpha$ we get following equations:
\begin{equation}
\begin{split}
2 \CG_0 +  \CG_0'^2 &= (1-\U^2 x^2),\\
2 \CG_1 + 2 \CG_1' \CG_0' &= -2,\\
2 \CG_2 + 2 \CG_0' \CG_2' +  \CG_1'^2 &= \frac{1}{1-\U^2 x^2},\\
2 \CG_3 + 2 \CG_0' \CG_3' + 2 \CG_1' \CG_2' &= 0,\\
2 \CG_4 + 2 \CG_0' \CG_4' + 2 \CG_1' \CG_3' + \CG_2'^2 &= 0,\\
\dots
\end{split}
\end{equation}
Where the primes denote the derivative w.r.t $x$.\\
The zeroth order equation is the equation for the uncharged rotating case, so the solution for $\CG_0$ should be the uncharged static black hole which we are perturbing.
\begin{equation}
2\CG_0 = 1 - C_{\pm} x^2,
\end{equation}
where
\begin{equation}
C_{\pm} = \frac{1\pm\sqrt{1-4\U^2}}{2}.
\end{equation}
Which gives $\CG_0'=-C_{\pm}~x$, and putting it in the next equation for $\CG_1$ gives
\begin{equation}\label{cbhg1}
\CG_1 = \kappa_1 x^{\frac{1}{C_{\pm}}} - 1.
\end{equation}
Let us pause  here to analyze the solution which gives two different conditions,\\\\
$\bf{1}.$ If $1/C{\pm}$ is  a not an even integer  then  after imposing regularity condition at $x=0$, we require $\kappa_1$ to vanish, giving $\CG_1 = -1$. Hence $\CG_1'=0$.\\\\
$\bf{2}.$If $1/C{\pm}$ is  an even integer  then homogeneous part becomes analytic at $x=0$, hence we cannot set the $\kappa_1$  to zero. \\\\
Here we will present only the solution when $1/C{\pm}$ is  a not an even integer.
The equation for $\CG_2$ becomes,
\begin{equation}
\CG_2 -C_{\pm}~x~\CG_2' = \frac{\frac{1}{2}}{1-\U^2 x^2}.
\end{equation}
Which gives
\begin{equation}
\CG_2 = \kappa_2 x^{\frac{1}{C_{\pm}}} - \frac{1}{2C_{\pm}}(\U x)^{\frac{1}{C_{\pm}}}\int_0^{\U x}\frac{y^{\frac{-1}{C_{\pm}}-1}}{1-y^2}dy.
\end{equation}
The particular integral part of the solution is analytic by itself as $x\rightarrow0$. This demands $\kappa_2 = 0$.
Hence
\begin{equation}
\CG_2 = - \frac{1}{2C_{\pm}}(\U x)^{\frac{1}{C_{\pm}}}\int_0^{\U x}\frac{y^{\frac{-1}{C_{\pm}}-1}}{1-y^2}dy
\end{equation}
\\The equation for $\CG_3$ is
\begin{equation}
\CG_3 -C_{\pm}~x~\CG_3' = 0,
\end{equation}
Which is a homogeneous equation, the solution of which we know is not analytic at $x\rightarrow 0$. So $\CG_3 = 0$.\\
And the equation for $\CG_4$ is
\begin{equation}
2\CG_4 - 2 C_{\pm}~x~\CG_4'= -\CG_2'^2.
\end{equation}
Here as well the particular integral turns out to be analytic, so we have to set the homogeneous solution to be zero.\\
\\Hence the shape function for a charged rotating membrane can be written perturbatively upto second order in $\alpha^2$ as
\begin{equation}\label{chargefin}
2\CG = (-C_{\pm}~x^2 + 1) - 2\alpha^2 + \left(- \frac{1}{2C_{\pm}}(\U x)^{\frac{1}{C_{\pm}}}\int_0^{\U x}\frac{y^{\frac{-1}{C_{\pm}}-1}}{1-y^2}dy\right)\alpha^4+\cdots
\end{equation}\\
\begin{figure}[!tbp]
		\centering
		\begin{minipage}[b]{0.48\textwidth}
		\includegraphics[width=\textwidth]{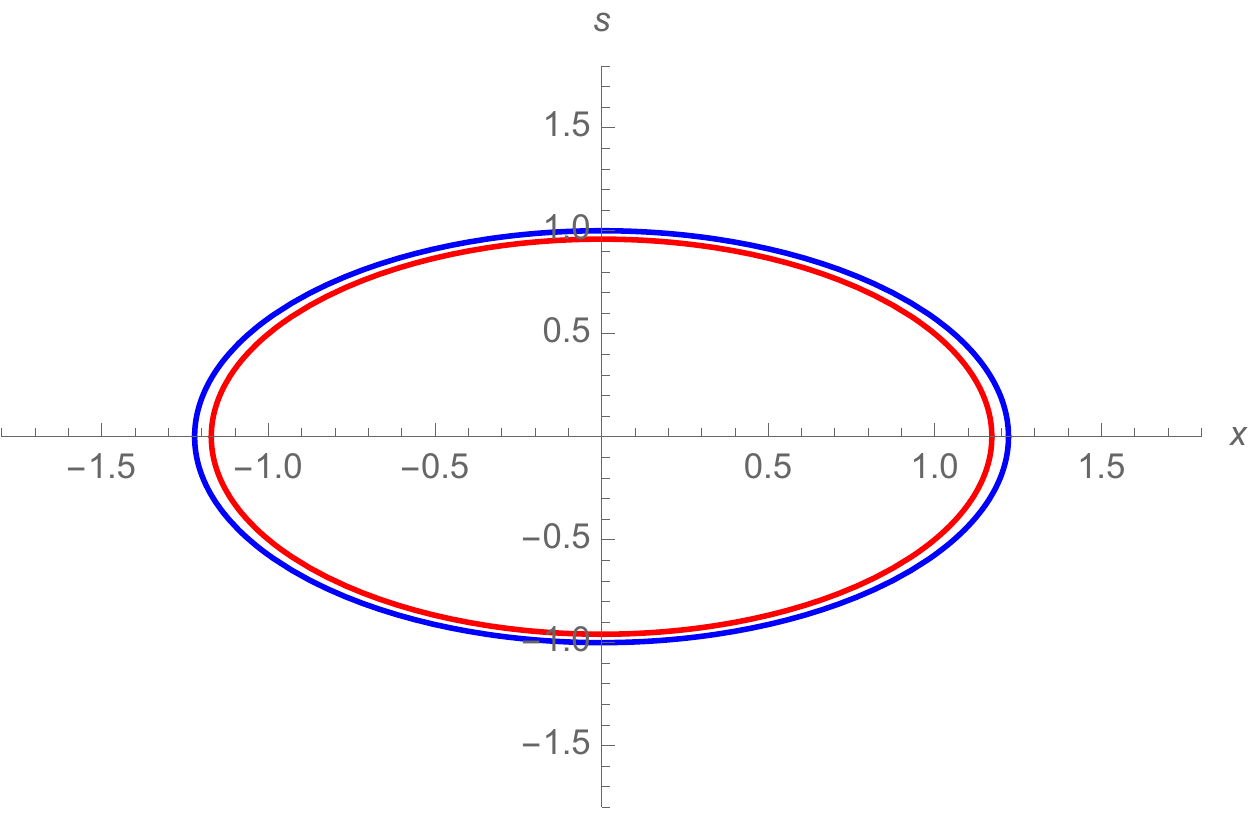}
		\caption{The inner red surface corresponds to the solution for the charged membrane with $\alpha=0.2$ and $\U=0.47$ for the $+$ branch while the outer blue surface is for the uncharged membrane.}
		\label{fig1}
	\end{minipage}
		\hfill
		\begin{minipage}[b]{0.48\textwidth}
		\includegraphics[width=\textwidth]{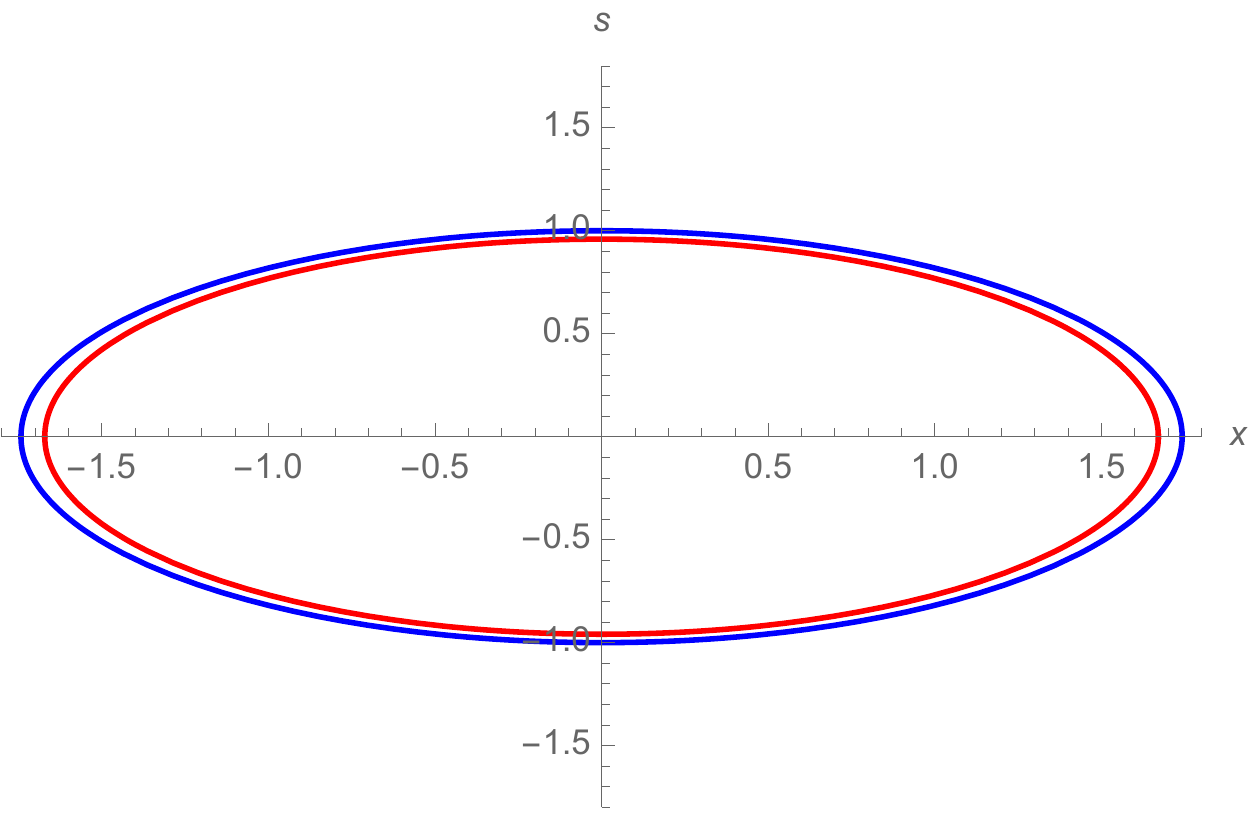}
		\caption{The inner red surface corresponds to the solution for the charged membrane with $\alpha=0.2$ and $\U=0.47$ for the $-$ branch while the outer blue surface is for the uncharged membrane.}
			\label{fig2}
		\end{minipage}	
\end{figure}
In figure \ref{fig1} and \ref{fig2} we show the surface comparison for the \eqref{chargefin} with the uncharged black hole solutions \eqref{actsol}. The inner red surface corresponds to the solutions for the charged membrane with $\alpha=0.2$ and $\U=0.47$. The charge membrane  is a deformed ellipsoid and it becomes  flatter compared  the uncharged rotating membrane for both the $+$ and $-$ branches.\\\newpage
{ \bf{Perturbatively adding rotation to static charged membrane}}\\\hspace{1cm}

In this section we will  solve the equation \eqref{efchfltnw}
for the simplest case where only one rotation is turned on and do a perturbative expansion in $\U^2$  about a static charged membrane. Let us assume that the expansion of the shape function in $\U^2$ is given by 
\begin{equation}
\label{expomchft}
\CG = \CG_0 + \U^2 \CG_1 + \U^4 \CG_2 + \U^6 \CG_3 + \dots
\end{equation}

Substituting in \eqref{efchfltnw} and separating out powers of $\U^2$ we get following equations:
\begin{equation}
\begin{split}
2\CG_0 + \CG_0'^2 &= (1-\alpha^2)^2,\\
2\CG_1 + 2\CG_1'g_0' &=  x^2(\alpha^4-1),\\
2\CG_2 + 2\CG_0'\CG_2' + \CG_1'^2 &= \alpha^4x^4,\\
2\CG_3 + 2\CG_0'\CG_3' + 2\CG_1'\CG_2' &= \alpha^4x^6,\\
2\CG_4 + 2\CG_0'\CG_4' + 2\CG_1'\CG_3' +\CG_2'^2 &= \alpha^4x^8,\\
\dots
\end{split}
\end{equation}
Where the primes denote the derivative w.r.t $x$.\\
The zeroth order equation is the equation for the charged static case, so the solution for $\CG_0$ should be the charged static membrane which we are perturbing.
\begin{equation}\label{CG0}
2\CG_0 = (1-\alpha^2)^2 - x^2.
\end{equation}
Which gives $\CG_0'=-x$.\\
If we look at the equation at any order, we see that the homogeneous part of all the equations is the same, hence any equation can be written as
\begin{equation}
2\CG_n - 2\CG_n' = f_n(x)
\end{equation}
Which has a simple solution
\begin{equation}
\CG_n = C_nx - \frac{x}{2}\int_0^x\frac{f_n(\rho)}{\rho^2}d\rho.
\end{equation}
For the first order we have $f_1(x) =  x^2(\alpha^4-1)$. So we get
\begin{equation}
\CG_1 = C_1 x- \frac{1}{2} x^2(\alpha^4-1).
\end{equation}
Since the solution has to be regular at $x = 0$, $C_1 = 0$.
Now at any order by induction we can show that the particular integral is a polynomial in $x^2$. So the only odd order term is the homogeneous part which has to be set to zero at each order for the sake of regularity at $x = 0$. i.e. $C_n = 0$ for all $n$.\\
For the second order $f_2(r) = \alpha^4 x^4 -(\alpha^4 - 1)^2x^2$. So we get
\begin{equation}
\CG_2 = -\frac{1}{6}\alpha^4 x^4 + \frac{1}{2}(\alpha^4 - 1)^2x^2.
\end{equation}
We can proceed in the similar way for higher orders in $\U$,
\begin{equation}
\begin{split}
\CG_3 =& -\frac{1}{10} x^6 \left(\alpha^ 4  \right)+\frac{2}{9} \alpha ^4 \left(\alpha ^4-1\right)  x^4-\left(\alpha ^4-1\right)^3   x^2,\\
\CG_4 =& -\frac{1}{14} x^8 \left(\alpha ^4  \right)+\frac{1}{225} \alpha ^4 \left(37 \alpha ^4-27\right)  x^6-\frac{14}{27} x^4 \left(\alpha ^4 \left(\alpha ^4-1\right)^2 \right)\\ &+\frac{5}{2} \left(\alpha ^4-1\right)^4 x^2,\\
\CG_5 =& -\frac{1}{18} x^{10} \left(\alpha ^4 \right)+\frac{2}{245} \alpha ^4 \left(17 \alpha ^4-10\right) x8-\frac{x^6 \left(\alpha ^4 \left(1471 \alpha ^8-2362 \alpha ^4+891\right) \right)}{3375}\\ &+\frac{116}{81} \alpha ^4 \left(\alpha ^4-1\right)^3 x^4-7 x^2 \left(\left(\alpha ^4-1\right)^5 \right),\\
\dots
\end{split}
\end{equation}
Plugging back the solutions in \eqref{expomchft} we get the solution for charged rotating membrane at the leading order in $1/D$ expansion.\footnote{As we go higher order in perturbations, the solutions for $\CG_n$ are polynomial of $x^{2n}$, which naively looks divergent at large values of $x$. However the radius of convergence of this perturbation series is $\frac{1}{|\U|}$ and perturbation series is well defined in this range. }

\subsection{The effective membrane equation for Stationary `axially-symmetric' configurations  in $AdS$ }\label{rotmemAdS}
In this subsection we will find axially-symmetric solutions to \eqref{Statsol} and \eqref{Statsolunch} in the ambient empty $AdS$  space both in presence and absence of charge  which will be dual to rotating black hole in presence and absence of charge respectively.\\
A time like killing vector field of \eqref{globadssplit}  which brings out the `axially-symmetric' configuration of the final spacetime and hence the final membrane configuration is given by

\begin{equation}
k=\frac{\partial}{\partial t}+\omega^{i}\frac{\partial}{\partial\theta_i}.
\end{equation}
Hence, the unit time-like velocity field of the membrane that we consider is given by
\begin{equation}\label{gammaads}
u^{\mu}=\gamma k^{\mu},\quad \text{where} \quad \gamma=\frac{1}{\sqrt{\left(1-\sum_{i}r_i^2 \omega_i^2\right)+\frac{1}{L^2}\sum_i (r_i^2+s^2)}}.
\end{equation}

We consider the global $AdS$ metric with manifest $SO(D-p-2)$ isometry \eqref{globalads1} and \eqref{globadssplit}. Since we view the membrane as propagating in an empty $AdS$ .
The extrinsic curvature is given by 
\begin{equation}\label{kappaads}
\begin{split}
K&=\frac{1}{\sqrt{-G}}n^s\partial_s(\sqrt{-G}),\\
K&=D\frac{n^s}{s}=D\frac{1+\frac{s^2}{L^2}-\frac{1}{L^2}\sum_i r_i \partial_i g}{\sqrt{s^2+\sum_i\left(\frac{\partial g}{\partial r_i}\right)^2+\frac{1}{L^2}\sum_i\left(s^2-r_i \partial_i g\right)^2}}\,\,.\\   
\end{split}
\end{equation}
\subsection{Rotating uncharged membrane solutions in  $AdS$ back ground}\label{rotsolAdS}
The most general solutions of the membrane in without   charge in a  stationary configurations is given by \eqref{Statsolunch}
\begin{equation}\label{effleqnads}
K=\frac{ \gamma}{\beta},
\end{equation}
where we have defined $\beta=\frac{D}{4\pi T}$.\\

Using the expressions of $\mathcal{K}$ and $\gamma $ from (\eqref{gammaads},\eqref{kappaads}) we get the effective equation for the surface defined on the $s^2=2g$ as
\begin{equation}\label{adsefeq1}
\begin{split}
& 2g+\sum_i(\partial_ig)^2+\frac{1}{L^2}\left(2g -\sum_{i}r_i\partial_ig\right)^2 \\
&=\beta^2 \left(1+2\frac{g}{L^2}-\frac{1}{L^2}\sum_ir_i\partial_ig\right)^2\left(1+\frac{2g}{L^2}-\sum_ir_i^2\left(\omega_i^2-\frac{1}{L^2}\right)\right).
\end{split}
\end{equation}
Let us redefine $g, r_i$ and $\omega_i$ such that we can rewrite \eqref{shapeflt} in an intrinsic scale independent form.
\begin{equation}
\label{redsurds}
g=\beta^2 \CG,\qquad r_i=\beta x_i ,\qquad L= \beta \CL\qquad \text{and} \qquad\omega_i=\frac{\U_i}{\beta},
\end{equation}
with this new redefinition \eqref{redsurds} we can rewrite \eqref{adsefeq1}
\begin{eqnarray}
\label{adsefeqnw}
&& 2\CG+\sum_i(\partial_i\CG)^2+\frac{1}{\CL^2}\left(2\CG -\sum_{i}x_i\partial_i\CG\right)^2\nonumber\\
&&= \left(1+2\frac{\CG}{\CL^2}-\frac{1}{\CL^2}\sum_ix_i\partial_i\CG\right)^2\left(1+\frac{2\CG}{\CL^2}-\sum_ix_i^2\left(\U_i^2-\frac{1}{\CL^2}\right)\right).
\end{eqnarray}

The above differential equation is relatively  hard to solve in general, however  as has been explained in the situation of the flat space rotating solution we can solve this by a general quadratic ansatz as follows. Let us turn on rotation only along one plane and the general ansatz
$$\CG(x)= \frac{A \CL^2}{2}-\frac{C }{2}(x-B\CL)^2,$$
where   $A,C$ and $B$ are functions of $\CL$. For a non-trivial solution to \eqref{adsefeqnw} we require that $C\neq0$. \\
Plugging the quadratic ansatz in \eqref{adsefeqnw} and equating the different powers of $x$ to zero we arrive at the following relations obeyed by the coefficients $A, B $  and $C$ are as follows
\begin{equation}
\label{algeqns}
\begin{split}
&\CL^2 \left(A-B^2 C\right) \left(A-B^2 C+1\right)-\left(A-B^2 C+1\right)^3+B^2 C^2 \CL^2=0,\\
&\frac{4 B C \left(A-B^2 C+1\right)^2}{\CL}-2 B C \CL \left(A-B^2 C-C+1\right)=0,\\
&C \left(B^4 C (6 C-1)+B^2 \left(C \left(\CL^2-7\right)+2\right)+(C-1) \CL^2+1\right)-1
+A^2 \left(C+\CL^2 \U ^2-1\right)\\&+A \left(C \left(B^2 \left(-7 C-2 \CL^2 \U ^2+2\right)+2\right)+2 \CL^2 \U ^2-2\right)+\CL^2 \U ^2 \left(B^2 C-1\right)^2=0,\\
&2 B C \left(\CL^2 \U ^2 \left(A+B^2 (-C)+1\right)+A C-A-2 B^2 C^2+B^2 C+C-1\right)=0,\\
&B^2 C^2 \left(C+\CL^2 \U ^2-1\right)=0.
\end{split}
\end{equation}
There are five equations with three  undetermined variables, which makes the system of equations over-constrained. The last equation in \eqref{algeqns}  gives two possible solutions
\begin{equation}\label{solbc}
\begin{split}
B&=0,\text{or}\\ 
C&=1-\CL ^2 \U^2.
\end{split}
\end{equation}
Now  if we consider the second solution in \eqref{solbc} the system of equations in \eqref{algeqns} is still over-constrained and doesn't have any solution, hence the only consistent solution is $B=0$. Setting $B=0$ gives us two equations in terms of two   undetermined variables. Hence the ansatz which gives a consistent solution is 
$$\CG(x)= \frac{A \CL^2}{2}-x^2\frac{C }{2}.$$
\begin{equation}
\label{algeqnscon}
\begin{split}
&\CL^2 \left(A^2+A\right)-(A+1)^3=0,\\
&(C-1) \left((A+1)^2+C \CL^2\right)+(A+1)^2 \CL^2 \U ^2=0.\\
\end{split}
\end{equation}
The first equation on \eqref{algeqnscon} can be solved and we have the following set of solutions 
\begin{equation}\label{solads1}
\begin{split}
A&= -1,\\
A_{\pm}&=\frac{1}{2} \left(\left(\CL^2\pm \CL \sqrt{\CL^2-4}\right)-2\right).\
\end{split}
\end{equation}
Plugging back the solutions for $A_i$ in the second equation in \eqref{algeqnscon} we  get a class of different solutions 
\begin{itemize}
	\item \begin{equation}\label{sol1}\begin{split}
	A&=-1,\\C&= 0 \qquad \text{or}\qquad 1.
	\end{split}  	\end{equation}
	\item \begin{equation}\label{sol2}\begin{split}
	A_{+}&=\frac{1}{2} \left(\left(\CL^2+ \CL \sqrt{\CL^2-4}\right)-2\right),\\
	C_{+\pm}&=-\frac{1}{2}\left(A_{+}-1\right)\pm\frac{1}{2}\CL\sqrt{A_{+}(1-4\U^2)}.
	\end{split}	\end{equation}
	\item \begin{equation}\label{sol3}\begin{split}
	A_{-}&=\frac{1}{2} \left(\left(\CL^2- \CL \sqrt{\CL^2-4}\right)-2\right),\\
	C_{-\pm}&=-\frac{1}{2}\left(A_{-}-1\right)\pm\frac{1}{2}\CL\sqrt{A_{-}(1-4\U^2)}.
	\end{split}	\end{equation}
\end{itemize}

The solution \eqref{sol1} gives a hyperbolic solution which is non-compact and hence an unphysical solution, hence we discard the solution. 
The only physical solutions to \eqref{adsefeq1} are \eqref{sol2} and \eqref{sol3} .

Let us now consider the validity regime of the solutions \eqref{sol2} and \eqref{sol3} , which gives us a compact and real solutions. For compactness of the solutions we require that \footnote{In this section and in the next where we obtain perturbative solutions for charged membranes, we consider only the compact solutions. There are valid noncompact solutions as well ,which are discussed in \ref{noncomp}}
\begin{equation}
\label{valsol}
\begin{split}
& A_{\pm}>0,\qquad C_{\pm\pm}>0,
\end{split}
\end{equation}
for the solutions to be real  we require \footnote{The other condition;i.e, $\CL<-2$ is excluded from the physical requirement that $\CL=\beta L$ and both the temperature and the $AdS$ radius scale are positive.}
\begin{equation}
\label{valsol2}
\begin{split}
& \CL \geq	2,\qquad -\frac{1}{2}<\U\leq	\frac{1}{2}.
\end{split}
\end{equation}

In terms of the thermodynamic quantities  we get a real and consistent rotating black hole (black brane) solutions in $AdS$ when the thermodynamic temperature $T$ and the rotation $\omega$ follows 
\begin{equation}
\begin{split}
T>\frac{D}{2\pi L} \quad\text{and} \qquad\frac{-2\pi T}{D}<\omega<\frac{2\pi T}{D}.
\end{split}
\end{equation}\\
$\CL\geq2$ also ensure that $A_{\pm}>0$ as required for  compactness of the solution.
The requirement of compactness of the  solutions to \eqref{adsefeqnw} gives the following conditions as listed in \ref{tab1}.
\begin{table}
	\begin{center}
		\begin{tabular}{|l|r|c|}\hline
			\hline
			Parameter range & allowed $A$ & allowed $C$\\
			\hline
			\hline
			$|\Upsilon| < \frac{1}{\mathcal{L}}$ & $A_+$ & $C_{++}$ \\
			& $A_-$ & $C_{-+}$ \\
			\hline
			$\frac{1}{\mathcal{L}} < |\Upsilon| \leq \frac{1}{2}$ & $A_-$ & $C_{-+}$ \\
			& $A_-$ & $C_{--}$\\
			\hline
		\end{tabular}
		\caption{Allowed branches of solutions for different values of $\U$}\label{tab1}
	\end{center}
\end{table}\\

\begin{figure}[!tbp]
	\centering
	\begin{minipage}[b]{0.47\textwidth}
		\includegraphics[width=\textwidth]{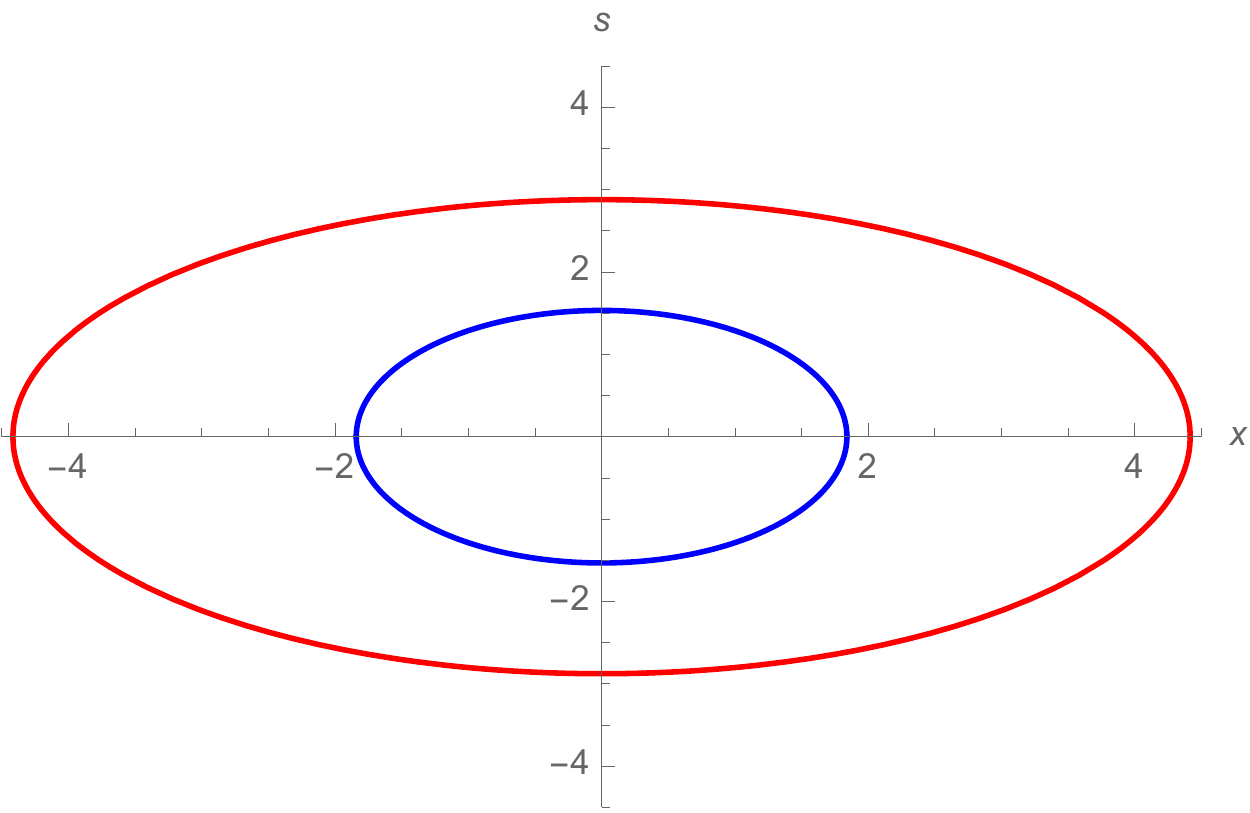}
		\caption{\textcolor{red}{$s^2+ r^2 C_{(++)}= A_{+} L^2$} and \textcolor{blue}{$s^2+ r^2 C_{(-+)}= A_{-} L^2$} with $\CL= 2.1, \U=0.4$}
		\label{figads1}
	\end{minipage}
	\hfill
	\begin{minipage}[b]{0.47\textwidth}
		\includegraphics[width=\textwidth]{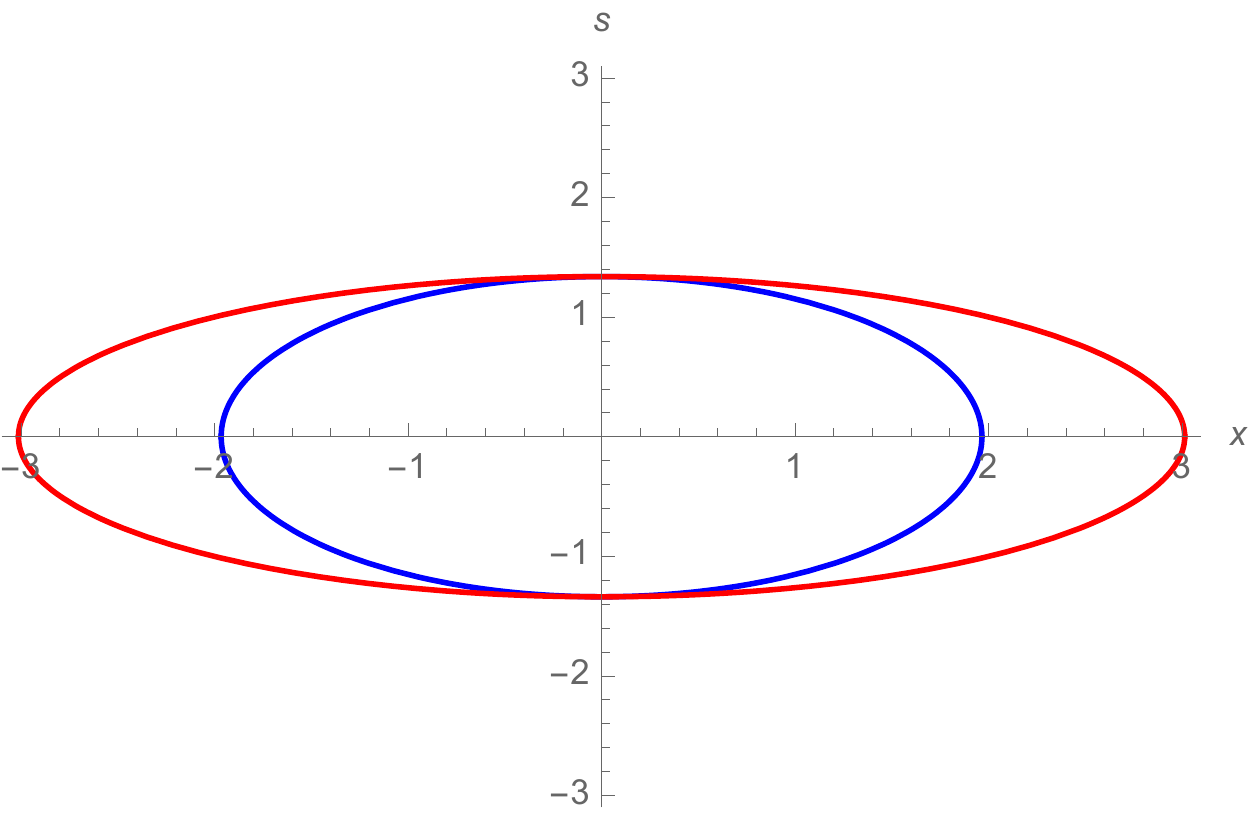}
		\caption{ \textcolor{red}{$s^2+ r^2 C_{(-+)}= A_{-} L^2$} and \textcolor{blue}{$s^2+ r^2 C_{(--)}= A_{-} L^2$} with $\CL= 2.3, \U=0.49$ }
		\label{figads2}
	\end{minipage}
\end{figure}

$C_{+-} $ is always negative  for any acceptable values of $\CL$ and $\U$ as given in \eqref{valsol2} and hence is never give a compact solution .\\
A detailed derivation of these limits is  presented in \ref{valsolap}.
Hence only $C_{++}$, $C_{-+}$ and $C_{--}$ give a real and compact solution for the surface.
The corresponding  membrane are given by
\begin{equation}\label{adstsol}\begin{split}
s^2+ r^2 C_{(++)}= A_{+} L^2,\\
s^2+ r^2 C_{(-+)}= A_{-} L^2,\\
s^2+ r^2 C_{(--)}= A_{-} L^2.\\
\end{split}
\end{equation}
These are  ellipsoidal membrane which are fatter in the plane of rotation  
all other directions (this excess bulge is a consequence of the 
centrifugal force). \\Figure \ref{figads1} shows the surface for turning on rotation in one  plane for the first entry in \ref{tab1} and figure \ref{figads2}  second entry in \ref{tab1} respectively. 

\newpage
\subsection{Rotating charged membrane solution in  $AdS$ }

The most general stationary charged membrane solutions in $AdS$ is given by 
\begin{equation}
\begin{split}
Q&=\alpha\gamma,\\
K&= \frac{\gamma}{\beta (1-Q^2)},
\end{split}
\end{equation}

where we have used the definition 
$$\alpha=2\sqrt{2\pi} \mu$$ and $$\beta=\frac{D}{4\pi T}.$$

The extrinsic curvature is given by \eqref{kappaads} and \eqref{gammaads} and the  effective equation for the surface is given by 

\begin{equation}
\begin{split}
\label{adsefeqch}
& \left(2g+\sum_i(\partial_ig)^2+\frac{1}{L^2}\left(2g -\sum_{i}r_i\partial_ig\right)^2\right)\left(1+\frac{2g}{L^2}-\sum_ir_i^2\left(\omega_i^2-\frac{1}{L^2}\right)\right)\\
&=\beta^2 \left(1+2\frac{g}{L^2}-\frac{1}{L^2}\sum_ir_i\partial_ig\right)^2\left(1+\frac{2g}{L^2}-\sum_ir_i^2\left(\omega_i^2-\frac{1}{L^2}\right)-\alpha^2\right)^2.
\end{split}
\end{equation}
With the redefinition \eqref{redsurds} we can rewrite \eqref{adsefeqch} in an intrinsically scale independent form as  
\begin{equation}\begin{split}
\label{adsefeqchnw}
& \left(2\CG+\sum_i(\partial_i\CG)^2+\frac{1}{\CL^2}\left(2\CG -\sum_{i}x_i\partial_i\CG\right)^2\right)\left(1+\frac{2\CG}{\CL^2}-\sum_ix_i^2\left(\U_i^2-\frac{1}{\CL^2}\right)\right)\nonumber\\
&= \left(1+2\frac{\CG}{\CL^2}-\frac{1}{\CL^2}\sum_ix_i\partial_i\CG\right)^2\left(1+\frac{2\CG}{\CL^2}-\sum_ix_i^2\left(\U_i^2-\frac{1}{\CL^2}\right)-\alpha^2\right)^2.
\end{split}
\end{equation}
\newpage
The above equation  is  hard to solve non-perturbatively. However we can solve it  perturbatively in two different scheme
\begin{itemize}
	\item Perturbatively adding charge to neutral rotating membrane
	\item Perturbatively adding rotation to static charged membrane
\end{itemize}
\vspace{1cm}
{\bf{Perturbatively adding charge to the neutral rotating membrane}}\\

We will consider the  simplest case where only one rotation is turned on. Let us expand the shape function perturbatively in terms of $\alpha^2$ as
\begin{equation}\label{expchg}
\CG = \CG _0 + \alpha^2 \CG _1 + \alpha^4 \CG_2 + \alpha^6 \CG_3 + \cdots
\end{equation}

Substituting \eqref{expchg} in \eqref{adsefeqchnw} and separating out powers of $\alpha$ we get the equations for the different orders in the expansions.\\
The zeroth order equation is 
\begin{equation}
\begin{split}
&\frac{x^2 \CG _0'(x)^2}{\CL^2}-\frac{4 x \CG _0(x) \CG _0'(x)}{\CL^2}+\frac{\left(-2 \CG _0(x)+\CL^2 \left(x^2 \U ^2-1\right)-x^2\right) \left(-x \CG _0'(x)+2 \CG _0(x)+\CL^2\right)^2}{\CL^6}\\&+\CG _0'(x)^2+\frac{4 \CG _0(x)^2}{\CL^2}+2 \CG _0(x)=0,\\
\end{split}
\end{equation}
Where the primes denote the derivative w.r.t $x$.\\
The zeroth order equation is the equation for the uncharged rotating case, so the solution for $\CG_0$ should be the uncharged rotating membrane .
\begin{equation}
2\CG_0 = A\CL^2- C x^2,
\end{equation}
where $A$ and $C$ are given in \eqref{sol1} and \eqref{sol2}. For each of the different allowed solutions, $A_{+}, C_{++}$ and $A_{-}, C_{-+},C_{--}$ we will have the different charged solutions in $AdS$. For simplicity we will denote the these by $A$ and $C$. \\
Substituting the   zeroth order solution in \eqref{adsefeqchnw} the equation for the first order in $\alpha^2$ is given by   
\begin{equation}\label{eqg1}
\begin{split}
\left(a x+b x^3\right) \CG_1'(x)+\CG_1(x) \left(a_1+b_1 x^2\right)+c=0,
\end{split}
\end{equation}
where the form of the coefficients are given by
\begin{equation}
\begin{split}
&a=\frac{(A+1)^2}{\CL^2}-A-C,\\
&b=-\frac{(A+1) \left(C+\CL^2 \U ^2-1\right)}{\CL^4},\\
&c=(A+1)^2,\\
&a_1=-\frac{3 (A+1)^2}{\CL^2}+2 A+1,\\
&b_1= \frac{2 (A+1) \left(C+\CL^2 \U ^2-1\right)}{\CL^4}.
\end{split}
\end{equation}
The solution to \eqref{eqg1}is given by 
\begin{equation}\label{cbhadsg1}
\CG_1(x)=\kappa  x^{-\frac{a_1}{a}} \left(a+b x^2\right)^{\frac{1}{2} \left(\frac{a_1}{a}-\frac{b_1}{b}\right)}-\frac{c \, _2F_1\left(1,\frac{b_1}{2 b};\frac{a_1}{2 a}+1;-\frac{b x^2}{a}\right)}{a_1}.
\end{equation}
After imposing regularity condition at $x=0$, we require $\kappa$ to vanish\footnote{Here we consider the case where $a_1/a$ is not an even integer}, giving $\CG_1 = \frac{c \, _2F_1\left(1,\frac{\text{b1}}{2 b};\frac{\text{a1}}{2 a}+1;-\frac{b x^2}{a}\right)}{\text{a1}}$.\\
Substituting the values of the coefficients $a,b,a_1,b_1$ and $c$  we can rewrite the solution for the $\CG_1(x)$ as 
\begin{equation}\label{solg1}
\begin{split}
\CG_1(x)&=\frac{x^2 \left(2 (A+1)^3 \left(C+\CL^2 \U ^2-1\right)\right)}{-(2 A+1) (2 C-1) \CL^4-2 (A+1)^2 \CL^2 (A-3 C+2)+3 (A+1)^4}\\&+\frac{(A+1)^2 \CL^2}{3 (A+1)^2-(2 A+1) \CL^2}.
\end{split}
\end{equation}
Substituting the expression  for the first order solution in \eqref{adsefeqchnw}, the equation for the  order  $\alpha^4$ is given by   
\begin{equation}\begin{split}\label{eqg2}
\CG_2'(x) \left(a_2 x^5+b_2 x^3+c_2 x\right)+\CG_2(x) \left(a_3 x^4+b_3 x^2+c_3 \right)+\left(a_4 x^4+b_4 x^2+c_4\right)=0,
\end{split}
\end{equation} where the coefficients $a_2,a_3,a_4,b_2,b_3,b_4,c_2,c_3$ and $c_4$ are given in appendix \ref{coeffads} 

We can write \eqref{eqg2} in a simple form as
\begin{equation}\begin{split}
&\CG_2'(x) +\CG_2(x) P(x)+q(x)=0, \qquad\text{where}\\
&P(x)=\frac{\left(a_3 x^4+b_3 x^2+c_3 \right)}{\left(a_2 x^5+b_2 x^3+c_2 x\right)},\qquad q(x)=\frac{\left(a_4 x^4+b_4 x^2+c_4\right)}{\left(a_2 x^5+b_2 x^3+c_2 x\right)}.
\end{split}
\end{equation}
The solution is of the form 
\begin{equation}\begin{split}\label{g2sol}
\CG_2(x)= -e^{-\int{P(x)dx}}\int_{0}^{x}\left(e^{\int{P(y)dy}}q(y)dy\right)+\kappa e^{-\int{P(x)dx}},
\end{split}
\end{equation}
regularity of the solution forces $\kappa$ to vanish.\\
Hence the shape function for a charged rotating black hole can be written perturbatively upto second  order in $\alpha^2$ is given by
\begin{equation}
2\CG = (A \CL^2 -Cx^2 \beta) + 2\alpha^2 \CG_1 + 2\CG_2\alpha^4+\cdots
\end{equation} where $\CG_1$ and $\CG_2$ are given in \eqref{solg1} and \eqref{g2sol} respectively.\\\vspace{1cm}

{\bf{Perturbatively adding rotation to the charged static  membrane}}\\

 Let us expand the shape function of the charged rotating membrane perturbatively in terms $\U^2$ as
\begin{equation}
\CG = \CG_0 + \U^2 \CG_1 + \U^4 \CG_2 + \U^6 \CG_3 + \dots
\end{equation}

Substituting in \eqref{adsefeqchnw} and separating out powers of $\U$ , the zeroth order equation:
\begin{equation} 
\begin{split}
& \frac{x^2 \CG_0'(x)^2}{\CL^2}-\frac{4 x \CG_0(x) \CG_0'(x)}{\CL^2}-\frac{\left(2 \CG_0(x)-\left(\alpha ^2-1\right) \CL^2+x^2\right)^2 \left(-x \CG_0'(x)+2 \CG_0(x)+\CL^2\right)^2}{\CL^6 \left(2 \CG_0(x)+\CL^2+x^2\right)}\\&+\CG_0'(x)^2+\frac{4 \CG_0(x)^2}{\CL^2}+2 \CG_0(x)=0.
\end{split}
\end{equation}
Where the primes denote the derivative w.r.t. $x$.\\
The solution to this equation is given by
\begin{equation}\label{g0sol}
2\CG_0 = {\cal{A}}_{\pm} - x^2,
\end{equation}
where ${\cal{A}}_{\pm}=\frac{1}{2} \left(\left(2 \alpha ^2\pm \CL \sqrt{4 \alpha ^2+\CL^2-4}\right)+\CL^2-2\right).$
Which gives $\CG_0'=-x$.\\
The equation at any order can be written in the form 

\begin{equation}
X(\alpha,\CL)\CG_n - Y(\alpha,\CL)\CG_n' =Z(\alpha,\CL) f_n(x).
\end{equation}
Which has a simple solution
\begin{equation}
\CG_n =C_n e^{\frac{x X}{Y}}+e^{\frac{x X}{Y}} \int_1^x -\frac{Z f_n(r) e^{-\frac{r X}{Y}}}{Y} \, dr.
\end{equation}
For the first order  we get\footnote{Here we have used the positive branch of the solution in \eqref{g0sol}, $$2\CG_0 = \frac{1}{2} \left(\left(2 \alpha ^2+\CL \sqrt{4 \alpha ^2+\CL^2-4}\right)+\CL^2-2\right) - x^2.$$ }
\begin{equation}\begin{split}
\CG_1 = &C_1 x^{2-\frac{\CL^3+\CL^2 \sqrt{4 \alpha ^2+\CL^2-4}+2 \alpha ^2 \sqrt{4 \alpha ^2+\CL^2-4}+4 \alpha ^2 \CL}{2 \CL}}\\+&\frac{x^2 \left(\CL^2 \left(\CL^3+\CL^2 \sqrt{4 \alpha ^2+\CL^2-4}+2 \alpha ^2 \sqrt{4 \alpha ^2+\CL^2-4}+\left(4 \alpha ^2-2\right)\CL\right)\right)}{2 \left(\CL^3+\CL^2 \sqrt{4 \alpha ^2+\CL^2-4}+2 \alpha ^2 \sqrt{4 \alpha ^2+\CL^2-4}+4 \alpha ^2 \CL\right)}.
\end{split}
\end{equation}
Since the solution has to be regular at $x = 0$, $C_1 = 0$.
We have checked upto order $\U^8$ and we find 
 that the particular integral is even order polynomial in $x$. So the only odd order term is the homogeneous part which has to be set to zero each order for the sake of regularity at $x = 0$. i.e. $C_n = 0$ for all $n$ for $n\leq4$.\\
For the second order  we get
\begin{equation}\label{solg2}
\CG_2 = \zeta_1(\alpha,\CL) x^2+\zeta_2 (\alpha,\CL)x^4,
\end{equation}
where  $\zeta_1$ and $\zeta_2$ are given in appendix \ref{coeffads}

The higher order solutions are given below, where the constants $\zeta_i's$ are functions of $\alpha$ and $\CL$ only,
\begin{equation}
\begin{split}\label{higsol}
\CG_3 =&  x^6 \zeta_3+\zeta_4 x^4+\zeta_5 x^2,\\
\CG_4 =&  x^8 \zeta_6+\zeta_7 x^6 +\zeta_8 x^4+\zeta_9 x^2,\\
\dots
\end{split}
\end{equation}
The coefficients  $\zeta_i's$ are quite complicated functions of  $\alpha$ and $\CL$, but the general structure of the higher order solutions are given by \eqref{higsol}, we don't present the explicit form of the coefficients here.\footnote{As we go higher order in perturbations, the solutions for $\CG_n$ are polynomial of $x^{2n}$, which naively looks divergent at large values of $x$. However the radius of convergence of this perturbation series is $\frac{1}{|\U|}$ and perturbation series is well defined in this range. }
\section{Thermodynamics for the rotating membrane solutions}\label{rottherm}
We have  derived the thermodynamic relations and the first law of thermodynamics in \ref{unch1stlaw} . In this section we  specialize these  relations for the rotating membrane configurations. 
We define a  conserved current  from the leading order stress tensor as
\begin{equation}
\begin{split}
J_{(E)}^\mu &= -T^{\mu\nu}k_\nu \\
&=(1+Q^2)\frac{K}{16\pi\gamma}u^\mu\left(\gamma^2-\sum_ir_i^2\omega_i^2\gamma^2\right).\\
\end{split}
\end{equation}
Now let's consider a spacelike slice of the membrane world volume with normal along $dt$. The conserved `total  energy' on this slice is given by

\begin{equation}
\label{toten}
\tilde{ E} = \int d^{D-2}y \sqrt{-G} J_{(E)}^0 = \int d^{D-2}y \sqrt{-G} (1+Q^2)\frac{K}{16\pi}\left(\gamma^2-\sum_ir_i^2\omega_i^2\gamma^2\right),
\end{equation}
since $u^0 = \gamma$. Note that this `total energy' also includes the contribution from angular momentum since  $k$ has contribution from the rotation Killing vectors.
We define the intrinsic energy as \newpage
\begin{equation}
\label{inen}
E  = \int  d^{D-2}y \sqrt{-G} (1+Q^2)\frac{\gamma^2K}{16\pi}.
\end{equation}
The angular momentum density is  simply the $T^{0\theta_i}$ component of the leading order stress-energy tensor defined on a spacelike slice on the membrane world volume with normal along $dt$.
\begin{equation}
T^{0\theta_i}=(1+Q^2)\frac{K}{16\pi} \delta(\rho-1)u^0u^{\theta_i},
\end{equation}
angular momentum  ${\cal J}_i$ in the $i'th$ plane is   defined as  
\begin{equation}
\label{anmom}
{\cal J}_i = \int d^{D-2}y \sqrt{-G}g_{\theta_i\theta_j}T^{0\theta_j} = \int d^{D-2}y \sqrt{-G}\sum_i\omega_ir_i^2\frac{\gamma^2(1+Q^2) K}{16\pi}.
\end{equation}
where $g_{\theta_i\theta_j}=r_i^2\delta_{ij}$ from \eqref{flsplit} and \eqref{globadssplit}.\\
Using \eqref{inen} and \eqref{anmom} we can rewrite \eqref{toten} as 
\begin{equation}
\tilde{E}= E-\sum_i\omega_i{\cal J}_i.
\end{equation}

Using \eqref{memcharge}, \eqref{mementropy} and 
\begin{equation}\label{varthermrot}
\delta   {\cal J}_i = \int d^{D-2}y  \delta\sqrt{-G}\sum_i\omega^ir_i^2\frac{\gamma^2(1+Q^2) K}{16\pi},
\end{equation}

we can simplify the combination $T\delta S ,\mu\delta q$ and $\delta J_i$ where $T$ and $\mu$ are defined in \eqref{Statsol}.
\begin{equation}\label{thermoproofrot}
\begin{split}
T\delta S +\mu\delta q +\sum_i\omega_i\delta  {\cal J}_i &= \int d^{D-2}y~ \delta\sqrt{-G}\left(T\frac{\gamma}{4} + \mu\frac{KQ\gamma}{2\sqrt{2\pi}}\right),\\
&= \int  d^{D-2}y~ \delta\sqrt{-G}\left(\frac{(1-Q^2)K}{4\pi\gamma}\frac{\gamma}{4} + \frac{Q}{2\sqrt{2\pi}\gamma}\frac{KQ\gamma}{2\sqrt{2\pi}}+\frac{1+Q^2}{16 \pi}K\gamma^2\right),\\
&= \int  d^{D-2}y~ \delta\sqrt{-G}~(1+Q^2)\frac{K}{16\pi}\gamma^2,\\
&= \delta E.\\
T\delta S+\mu \delta q&=\delta E-\sum_i\omega_i \delta  {\cal J}_i .
\end{split}
\end{equation}

Which is  the statement that rotating  membrane configurations  obey the first law of thermodynamics at the leading order in large $D$.\newpage

\subsection{Explicit form of the thermodynamic quantities for rotating membranes}

\hspace{1cm} Now let's calculate at the leading order in $D$ the thermodynamic quantities associated with the membrane rotating in a single plane. From equations \eqref{memtoten}, \eqref{memcharge}, \eqref{mementropy}, \eqref{inen} and \eqref{anmom} we notice that the densities of all the thermodynamic quantities on the membrane are functions of $r^2$. let $f(r^2)$ be the density for a quantity $F$. Therefore

\begin{equation}
\begin{split}
F &= \int dr d\theta d\Omega_{D-4} \sqrt{-G} f(r^2),\\
&= \int dr \left(\int d\theta d\Omega_{D-4} \sqrt{-G} \right) f(r^2),\\
&= 2\pi \Omega_{D-4} \int dr~r \left(2g(r^2)\right)^{\frac{D-4}2} f(r^2).
\end{split}
\end{equation}

We have used the fact that for both the flat and AdS spaces in our choice of coordinates 
$$\int d\theta d\Omega_{D-4} \sqrt{-G} = 2\pi r \Omega_{D-4} s^{D-4} = 2\pi \Omega_{D-4} r \left(2g(r^2)\right)^{\frac{D-4}2}.$$

Due to a steep dependence of the integrand on $r$, we can utilize the saddle point approximation to do this integration. To obtain the saddle point we extremist the integrand w.r.t. $r$. Let the saddle point be $(r_*)^2 = \epsilon_0 + \frac{\epsilon_1}{D-4}$. Now we proceed to finding the saddle point

\begin{equation}
 \left((D-4) (r_*)^2 \frac{2g'((r_*)^2)} {2g((r_*)^2)} + 1\right)f((r_*)^2) + (r_*)^2f'((r_*)^2) = 0.
\end{equation}
Solving at the leading order in $\frac{1}{D-4}$, we get $\epsilon_0 = 0$ while at the subleading order we get $\epsilon_1 = -\frac{2g'(0)} {2g(0)}$ which is positive as $2g$ is a monotonically decreasing function of $r^2$. This means
\begin{equation}
(r_*)^2 = -\frac{2g(0)} {2g'(0)(D-4)}.
\end{equation}
A couple of things are easily noticeable. First, the saddle point at the leading order is ${\cal{O}}(\frac{1}{D})$, so it's sufficient to calculate the quantities at the leading order in $r^2$ at the saddle point, because any higher order terms will be suppressed in $1/D$. And secondly, the leading order saddle point in independent of the density $f$ that is being integrated, so all the integrals here have the same saddle point.\\

Now in its general form $2g(r^2) = A - Cr^2 + \mathcal{O}(r^4)$, so $(r_*)^2 = \frac{A}{C(D-4)}$. Thus 
$$2g((r_*)^2) = A\left(1 - \frac{1}{D-4}\right) + \mathcal{O}\left(\frac{1}{(D-4)^2}\right).$$

Hence
\begin{equation}
2g((r_*)^2)^{\frac{D-4}2} = A^{\frac{D-4}2}\sqrt{e} \sim A^{\frac{D-4}2}.
\end{equation}
(Since we have omitted other constant factors coming from the saddle point integration, we can also omit other constant multiplicative factors like $\sqrt{e}$ consistently.)\\

Now we can write the expressions after the saddle point integration
\begin{equation}\label{thermsaddle}
\begin{split}
\tilde{E} &= V(1+\bar{Q}^2)\frac{\bar{K}}{16\pi},\\
E &= V(1+\bar{Q}^2)\frac{\bar{\gamma}^2\bar{K}}{16\pi},\\
\mathcal{J} &= V(1+\bar{Q}^2)\omega(r_*)^2\frac{\bar{\gamma}^2\bar{K}}{16\pi},\\
q &= V\frac{\bar{K}\bar{Q}\bar{\gamma}}{2\sqrt{2\pi}},\\
S &= V\frac{\bar{\gamma}}4,
\end{split}
\end{equation}
where $V = 2\pi\Omega_{D-4}\frac{A^{\frac{D-3}2}}{\sqrt{(D-4)C}}$ multiplied by all the consistently omitted factors like $\sqrt{e}$ and the ones from the saddle point integration. The barred quantities are the quantities evaluated at $r = r_*$. As all of $K$, $Q$ and $\gamma$ are nonzero at $r = 0$, the barred quantities can be computed at $r = 0$ instead of $r_*$.\\

Immediately we can see that $\mathcal{J}$ is subleading compared to $E$ because of the extra $(r_*)^2$ factor, which indicates that at the leading order in $D$ the contribution from the angular momentum to the total energy is negligible. Hence $\tilde{E} = E$ at the leading order. Also, $S$ doesn't have an explicit factor of $K$ in contrast to other quantities, it looks to be subleading as well. However it enters thermodynamics with a factor of $T$ multiplying it which itself is order $D$, so the entropy contributes to the thermodynamics in this way.\\

When the variations of these thermodynamic quantities are taken, the leading contribution comes from the variation of the `membrane volume' $V \sim A^{\frac{D-3}2}$ as expected. Thus we can infer that these thermodynamic quantities satisfy the relation 
\begin{equation}\label{thermrel}
\tilde{E} = \mu Q + T S.
\end{equation}
Now we calculate the thermodynamic quantities \eqref{thermsaddle} explicitly for a flat space charged rotating membrane. The barred quantities in \eqref{thermsaddle} are calculated at $r = 0$.  Also since $2g(r)$ is an even function of $r$, $g'(0) = 0$ and hence \eqref{shapeflt} gives
$$2g(0) = \beta^2(1-\alpha^2)^2,$$
where $\alpha = 2\sqrt{2\pi}\mu$ and $\beta = \frac{D}{4\pi T}$. Hence equations \eqref{kappa} and \eqref{gamma} give
\begin{equation}
\begin{split}
\bar{K} &= K(0) = \frac{D}{\sqrt{2g(0)}} = \frac{D}{\beta(1-\alpha^2)},\\
\bar{\gamma} &= \gamma(0) = 1,
\end{split}
\end{equation}
while \eqref{Qgam} leads to
\begin{equation}
\bar{Q} = \alpha\gamma(0) = \alpha.
\end{equation}
Plugging in \eqref{thermsaddle},
\begin{equation}\label{thermsaddle2}
\begin{split}
\tilde{E} &= V\frac{D(1+\alpha^2)}{16\pi\beta(1-\alpha^2)},\\
q &= V\frac{D\alpha}{2\sqrt{2\pi}\beta(1-\alpha^2)},\\
S &= \frac{V}4,
\end{split}
\end{equation}
which satisfy \eqref{thermrel}\\

Notice that the thermodynamic quantities depend on $\omega$ only through $V$, because $V$ depends on $C$ which in turn depends on $\omega$. Also, the angular momentum doesn't contribute as discussed earlier. So all the thermodynamic quantities associated with a charged rotating membrane are proportional to the corresponding quantities for the charged static spherical membrane at the same temperature. For a charged static spherical membrane, $C=1$ (see \eqref{CG0}) which means $\frac{V}{V_0} = \frac{1}{\sqrt{C}}$, where $V_0$ is the `membrane volume' of the charged spherical static membrane. So now \eqref{thermsaddle2} can be written as
$$\tilde{E} = \frac{\tilde{E}_0}{\sqrt{C}},~~q = \frac{q_0}{\sqrt{C}},~~S = \frac{S_0}{\sqrt{C}},~~T = T_0,~~\mu = \mu_0,$$
where the quantities with subscript $0$ belong to the static spherical membrane. \footnote{A similar analysis can be done for the AdS membrane and the thermodynamic quantities can be obtained, but from the physics point of view it brings nothing to the table other than what the analysis of the flat case illustrates. So we opt not to present those results here.}\newpage

\section{Discussion}

\hspace{1cm} The important results obtained in this paper can be summarized as follows:

\begin{itemize}
	\item The membrane equations in \cite{Bhattacharyya:2018ads} are specialized to the stationary membrane configurations, which yield two scalar equations which we call the `stationary membrane equations' (SMEs).
	\item The thermodynamic quantities namely energy, entropy and electric charge associated with a stationary membrane are read off from the effective stress tensor and charge current on that membrane, and it is shown that they satisfy the first law of thermodynamics if and only if the membrane satisfies the SMEs. The integration constants in the SMEs are identified with the temperature and chemical potential of the membrane.
	\item Exact solutions of the uncharged version of the SMEs are obtained in axially symmetric (rotating) case, and for the charged rotating stationary membranes perturbative solutions are obtained in both the flat background and the AdS background.
\end{itemize}

However in this paper we have not discussed the dynamical stability of these solutions. In many cases the rotating black holes are unstable and would change into a more stable configuration, see \cite{Dias:2010eu,Emparan:2003sy}. This feature is likely to be seen in large number of dimensions as well, and \cite{Emparan:2014jca} discusses the stability of uncharged rotating black holes in large $D$. We would like to do a stability analysis by checking the quasinormal modes about the charged rotating solutions obtained in this paper, but we leave it for future work.\\

The perturbative solution for a charged rotating membrane with charge added perturbatively to a rotating uncharged membrane as done in \eqref{cbhg1},\eqref{cbhadsg1} has a subtlety. The perturbative corrections yield homogeneous solutions at every order, but they can be fixed by demanding the solutions to be regular at $r = 0$ except in the case where $\frac{1}{C}$ is an even integer. In such a case we have not been able to figure out the mechanism by which the homogeneous solutions can be fixed, but we guess that the solutions wouldn't be finite in extent in $r$ unless these integration constants are set to zero. This may also be related to the perturbative instability of the rotating solutions in large $D$ reported in \cite{Tanabe:2016opw}.\\

In this paper we haven't considered the rotating membranes in the background with positive cosmological constant, i.e. de Sitter spacetime. Naively one can work with quadratic ansatz for membrane solution in dS background as well, after the analytic continuation $L^2 \rightarrow -L^2$ in sections \ref{rotmemAdS} and \ref{rotsolAdS}. One obvious consequence is that there is an extra solution compared to AdS case, namely $A = -1, C = 1$ in equation \eqref{sol1} which we throw away in the AdS case as doesn't give a valid solution. This solution is peculiar as it is independent of $\omega$, However the dS case deserves an analysis more careful than just a naive analytic continuation in order to get more interesting solutions and to avoid potential subtleties.\\

Finally, according to the particular membrane paradigm we are working in, the large $D$ membranes should have a one-to-one correspondence with black holes in large $D$. It would be interesting to obtain the metric and gauge field outside the black holes corresponding to the membrane solutions obtained in this paper by putting the explicit expressions for the shape function, charge and velocity fields in the expressions for metric and gauge field in \cite{Bhattacharyya:2018ads}.  
\vspace{2cm}

\section*{Acknowledgments}
\hspace{1cm} We are grateful to S. Minwalla for suggesting this problem to us and for being part of the original collaboration, and S. Bhattacharyya, S. Kundu and  P. Nandi for providing us with the draft of \cite{Bhattacharyya:2018ads} for using the charged membrane equations in AdS which are pivotal in our work. We would especially like to thank  S. Bhattacharyya,  A. O'Bannon, Y. Dandekar, A.Saha and Song He for several extremely useful discussions over the course of this project. We would also like to thank S. Bhattacharyya, S. Minwalla and A.Saha  and  for comments on a preliminary draft of this manuscript. The work of ST is supported by ITP-CAS. The work of MM was supported in part by an ISF excellence center grant 1989/14, a BSF grant 2016324 and  by the Israel Science Foundation under grant 504/13. 
\newpage

\appendix

\section{Identities for section \ref{subsecstat}}\label{idens2}

\hspace{1cm} To make the equations easy to the eye, we introduce a colour coding to various terms in many equations in this paper, especially in the first two appendices and in section \ref{EffcurrST}. We have maintained the leading order in all these equations to be ${\cal O}(D)$, and so whenever some manipulation gives subleading terms, they will be coloured \textcolor{blue}{blue} as soon as they appear, and then will disappear in the nest step, and lumped together with other subleading terms in the form of the ${\cal O}(1)$ symbol at the end of the line. 

The Gauss' equation for a timelike hypersurface

\begin{equation}\label{Gauss}
\begin{split}
\hat{R}_{\mu\nu\alpha\beta} &= R_{MNPQ}e^M_\mu e^N_\nu e^P_\alpha e^Q_\beta +\left(K_{\mu\alpha}K_{\nu\beta}- K_{\nu\alpha}K_{\mu\beta}\right),\\
\hat{R}_{\mu\nu} &= R_{MN} e^M_\mu e^N_\nu +\left(K~K_{\mu\nu}- K_\mu^\alpha K_{\nu\alpha}\right).
\end{split}
\end{equation}

\begin{equation}\label{gamp1}
(u\cdot\hat{\nabla})\gamma = \gamma^4k^{\mu}k^{\alpha}\hat{\nabla}_{\mu}k_{\alpha} =\frac{\gamma^4}2k^{\mu}k^{\alpha}(\hat{\nabla}_{\mu}k_{\alpha}+\hat{\nabla}_{\alpha}k_{\mu})= 0.
\end{equation}
due to the Killing equation.\\

\begin{equation}\label{gamp2}
\begin{split}
u\cdot\hat{\nabla} u_\mu &= u\cdot\hat{\nabla}(\gamma k_\mu),\\
&= \gamma u\cdot\hat{\nabla}k_\mu,\\
&= \gamma^2 k^\nu\hat{\nabla}_\nu k_\mu,\\
&= -\gamma^2 k^\nu\hat{\nabla}_\mu k_\nu,\\
&= -\frac{\gamma^2}2 \hat{\nabla}_\mu(k_\nu k^\nu),\\
&= \frac{\gamma^2}2 \hat{\nabla}_\mu(\gamma^{-2}),\\
&= -\hat{\nabla}_\mu\ln\gamma.
\end{split}
\end{equation}

\begin{equation}\label{uKupuRu}
\begin{split}
K~u\cdot K\cdot u + u\cdot R\cdot u &= u\cdot \hat{R}\cdot u ~\textcolor{blue}{+~u\cdot K\cdot K\cdot u},\\
&= -u^\nu\left(\hat{\nabla}_\mu\hat{\nabla}_\nu u^\mu \textcolor{blue}{- \hat{\nabla}_\nu\hat{\nabla}_\mu u^\mu}\right)+\mathcal{O}(1),\\
&= \hat{\nabla}^\mu\left(u\cdot\hat{\nabla}u_\mu\right) \textcolor{blue}{-(\hat{\nabla}^\mu u^\nu)(\hat{\nabla}_\nu u_\mu)}+\mathcal{O}(1),\\
&= -\hat{\nabla}^2\ln\gamma +\mathcal{O}(1)
\end{split}
\end{equation}
\eqref{gamp2} has been used here.

\begin{equation}\label{tnabsqu}
\begin{split}
p^\nu_\mu\hat{\nabla}^2 u_\nu &= p^\nu_\mu\hat{\nabla}^\alpha\hat{\nabla}_\alpha u_\nu\\
&= p^\nu_\mu\left(\hat{\nabla}^\alpha(k_\alpha\hat{\nabla}_\nu\gamma+k_\nu\hat{\nabla}_\alpha\gamma)-\hat{\nabla}_\alpha\hat{\nabla}_\nu u^\alpha\right)\\
&= \textcolor{blue}{p^\nu_\mu\left((\hat{\nabla}\cdot k) \hat{\nabla}_\nu\gamma + (k\cdot\hat{\nabla})\hat{\nabla}_\nu\gamma + (\hat{\nabla}^\alpha k_\nu)\hat{\nabla}_\alpha\gamma\right)} - p^\nu_\mu\left(\textcolor{blue}{\hat{\nabla}_\nu(\hat{\nabla}\cdot k)} + \hat{R}_{\nu\alpha}u^{\alpha}\right),\\
&= -p^\nu_\mu\hat{R}_{\nu\alpha}u^{\alpha} + \mathcal{O}(1).
\end{split}
\end{equation}

Therefore,
\begin{equation}\label{tnabsqupuK}
\begin{split}
p^\nu_\mu\left(\hat{\nabla}^2 u_\nu + K~u\cdot K_\nu\right) &= p^\nu_\mu\left(K~K_{\nu\alpha}-\hat{R}_{\nu\alpha}u^{\alpha}\right)+ \mathcal{O}(1),\\
&= -p^\nu_\mu e^M_\nu R_{MN} e^N_\alpha u^\alpha \textcolor{blue}{~+p^\nu_\mu (u\cdot K\cdot K)_\nu}+\mathcal{O}(1),\\
&= e^M_\mu\mathcal{P}_M^A R_{AN}u^N + \mathcal{O}(1),\\
&= \mathcal{O}(1)
\end{split}
\end{equation}

\begin{equation}\label{vmanip}
\begin{split}
\frac{(1+Q^2)}{(1-Q^2)}\hat\nabla_{\mu}\ln\gamma &= \frac{(1+\alpha^2\gamma^2)}{(1-\alpha^2\gamma^2)}\frac{\hat\nabla_{\mu}\gamma}\gamma\\
&= \frac{\frac{1}{\gamma^2}+\alpha^2}{\frac{1}{\gamma}-\alpha^2\gamma}\hat\nabla_{\mu}\gamma\\
&= -\hat\nabla_{\mu}\ln\left(\frac{1}{\gamma}-\alpha^2\gamma\right)
\end{split}
\end{equation}

\section{Identities for Section \ref{EffcurrST}}\label{idens3}

\begin{equation*}
\begin{split}
\hat{\nabla}^\mu \left(\frac{\hat{\nabla}^2u_\mu}{K}\right) &= \frac{\hat{\nabla}^\mu\hat{\nabla}^2u_\mu}{K} \textcolor{blue}{+\frac{(\hat{\nabla}^\mu K)\hat{\nabla}^2u_\mu}{K^2}},\\
&= \frac{\hat{\nabla}^\nu\hat{\nabla}_\mu\hat{\nabla}_\nu u^\mu}{K} + {\cal O}(1),\\
&= \textcolor{blue}{\frac{\hat{\nabla}^\nu\hat{\nabla}_\nu\hat{\nabla}_\mu u^\mu}{K}} + \frac{\hat{\nabla}^\nu(\hat{R}_{\mu\nu} u^\mu)}{K} + {\cal O}(1),\\
&= \textcolor{blue}{\frac{\hat{\nabla}^\nu(e^N_\nu (R\cdot u)_N)}{K}} + \frac{\hat{\nabla}^\nu(K~K_{\mu\nu}u^\mu)}{K} - \textcolor{blue}{\frac{\hat{\nabla}^\nu(K_\nu^\alpha K_{\alpha\mu} u^\mu)}{K}} + {\cal O}(1),\\
&= \hat{\nabla}^\nu(K_{\mu\nu}u^\mu) - \textcolor{blue}{\frac{(\hat{\nabla}^\nu K)K_{\mu\nu}u^\mu}{K}} + {\cal O}(1).
\end{split}
\end{equation*}

Hence
\begin{equation}\label{Bid1}
\hat{\nabla}^\mu \left(\frac{\hat{\nabla}^2u_\mu}{K} - (K\cdot u)_\mu\right) = {\cal O}(1).
\end{equation}

\begin{equation}\label{Bid2}
\begin{split}
\hat{\nabla}^\mu (u\cdot\hat{\nabla}u_\mu) &= u^\nu\hat{\nabla}_\mu\hat{\nabla}_\nu u^\mu + \textcolor{blue}{(\hat{\nabla}_\nu u^\mu)(\hat{\nabla}_\mu u^\nu)}\\
&= \textcolor{blue}{u^\nu\hat{\nabla}_\nu\hat{\nabla}_\mu u^\mu} + u^\nu\hat{R}_{\mu\nu} u^\mu + {\cal O}(1),\\
&= u\cdot R \cdot u + K~u\cdot K\cdot u - \textcolor{blue}{u\cdot K\cdot K\cdot u} + {\cal O}(1),\\
&= u\cdot R \cdot u + K~u\cdot K\cdot u + {\cal O}(1).
\end{split}
\end{equation}

\begin{equation}
\tilde{\Sigma}^\alpha_\alpha = \textcolor{blue}{\hat{\nabla}\cdot u} = {\cal O} \left(\frac{1}D\right)
\end{equation}
and
\begin{equation}
\tilde{\Sigma}_{\mu\nu} = \frac{1}2\left(p^\alpha_\mu\hat{\nabla}_\alpha u_\nu + p^\alpha_\nu\hat{\nabla}_\alpha u_\mu\right)
\end{equation}
also,
\begin{equation}
\begin{split}
\hat{\nabla}^\mu p^\alpha_\mu &=  \textcolor{blue}{(\hat{\nabla}\cdot u) u^\alpha + u\cdot \hat{\nabla}u^\alpha} = {\cal O}(1),\\
\hat{\nabla}^\mu p^\alpha_\nu &=  \textcolor{blue}{(\hat{\nabla}_\mu u_\nu) u^\alpha + u_\nu \hat{\nabla}_\mu u^\alpha} = {\cal O}(1)
\end{split}
\end{equation}
and
\begin{equation}
\begin{split}
u^\mu\hat{\nabla}^2u_\mu &= \hat{\nabla}^\nu(u^\mu\hat\nabla_\nu u_\mu) \textcolor{blue}{-(\hat{\nabla}^\nu u^\mu)(\hat{\nabla}_\nu u_\mu)} \\
&= {\cal O}(1)
\end{split}
\end{equation}
which means
\begin{equation}
\hat{\nabla}^2u_\mu = p^\nu_\mu\hat{\nabla}^2u_\nu + {\cal O}(1)
\end{equation}

Hence
\begin{equation}
\Sigma_{\mu\nu} = \tilde{\Sigma}_{\mu\nu} + {\cal O} \left(\frac{1}D\right)
\end{equation}
Thus
\begin{equation}\label{Bid3}
\begin{split}
\hat{\nabla}^\mu\Sigma_{\mu\nu} =& \hat{\nabla}^\mu\tilde{\Sigma}_{\mu\nu} + {\cal O} (1)\\
=& \frac{1}2\left(p^\alpha_\mu\hat{\nabla}^\mu \hat{\nabla}_\alpha u_\nu \textcolor{blue}{+(\hat{\nabla}^\mu p^\alpha_\mu)\hat{\nabla}_\alpha u_\nu}+ p^\alpha_\nu\hat{\nabla}^\mu\hat{\nabla}^\alpha u_\mu \textcolor{blue}{ + (\hat{\nabla}^\mu p^\alpha_\nu)\hat{\nabla}_\alpha u_\mu}\right)+ {\cal O} (1)\\
=& \frac{1}2\left(\hat{\nabla}^2u_\nu \textcolor{blue}{+u^\alpha u_\mu\hat{\nabla}^\mu \hat{\nabla}_\alpha u_\nu + p^\alpha_\nu\hat{\nabla}_\alpha\hat{\nabla}_\mu u^\mu}+ p^\alpha_\nu\hat{R}_{\mu\alpha} u^\mu \right)+{\cal O} (1)\\
=& \frac{1}2\left(\hat{\nabla}^2u_\alpha p^\alpha_\nu + K~p^\alpha_\nu K_{\mu\alpha} u^\mu \textcolor{blue}{-p^\alpha_\nu K_{\alpha}^{\beta} K_{\beta\mu} u^\mu}  \right)+{\cal O} (1)\\
=& \frac{1}2\left(\hat{\nabla}^2u_\alpha + K~ (K\cdot u)_\alpha \right)p^\alpha_\nu+{\cal O} (1)
\end{split}
\end{equation}

And finally
\begin{equation}\label{Vid}
\begin{split}
\hat{\nabla}\cdot{\cal V} =& \hat{\nabla}^\mu\left(\nabla_\mu Q + Qu\cdot\hat{\nabla}u_\mu\right)\\
=& \hat{\nabla}^2Q +Q\hat{\nabla}^\mu(u\cdot\hat{\nabla}u_\mu)\textcolor{blue}{+Q(\hat{\nabla}^\mu Q)(u\cdot\hat{\nabla}u_\mu)}\\
=&  \hat{\nabla}^2Q + Qu\cdot R\cdot u + QK~u\cdot K\cdot u+{\cal O}(1)
\end{split}
\end{equation}

\section{Flat space metric split in two planes}
The flat space metric in $D$ dimension can be written as
\begin{equation}\label{fltmet1}
ds^2=-dt^2+dr^2+r^2d\Omega^2_{D-2}.
\end{equation}

Now, we split the spacetime into $p$ two planes keeping a large $SO(D-2p-2)$ isometry intact by writing $$r^2=\sum_{i=1}^{2p}r_i^2+s^2.$$
\begin{equation}\label{flsplit}
ds^2=-dt^2+\sum_i(dr_i^2+r_i^2 d\theta_i^2)+ds^2+s^2 d\Omega^2_{D-2p-2}.
\end{equation}

\section{The Global AdS metric with manifest $SO(D-2p-2)$ isometry}
The metric of AdS written in global coordinates is given by
\begin{equation}\label{globalads1}
ds^2=-\left(1+\frac{r^2}{L^2}\right)dt^2+\frac{dr^2}{\left(1+\frac{r^2}{L^2}\right)}+r^2d\Omega^2_{D-2}.
\end{equation}

Now, we split the spacetime into $p$ two planes keeping a large $SO(D-2p-2)$ isometry intact by writing $$r^2=\sum_{i=1}^{2p}r_i^2+s^2.$$ Doing this the global AdS metric becomes
\begin{equation}\label{globadssplit}
ds^2=-dt^2\left(1+\frac{\sum {r_i}^2+s^2}{L^2}\right)-\frac{\left(\sum r_i  dr_i+s ds\right)^2}{\sum {r_k}^2+s^2+L^2}+\sum_i(dr_i^2+r_i^2 d\theta_i^2)+ds^2+s^2 d\Omega^2_{D-2p-2}.
\end{equation}

\section{All the uncharged rotating membrane solutions in flat and AdS backgrounds}

\hspace{1cm} It is important to note that the equations \eqref{shapeflt} and \eqref{adsefeq1} are actually the square of \eqref{effleqn} and \eqref{effleqnads} respectively. So solving them may give some spurious solutions which we need to weed out carefully. In this appendix we will obtain the conditions for a solution not to be spurious.\\

The large $D$ membrane paradigm is based on the following peculiarity of the large $D$ black holes that we consider. If we zoom into a patch around a point $y_0$ on the horizon of such a black hole, it looks like the spacetime near the horizon of a static black hole boosted with velocity $u(y_0)$, where $u$ is the velocity field of the membrane corresponding to that black hole. In other words, the membrane looks locally static at $y_0$ when seen from a frame of reference with a boost $u(y_0)$. Hence $u(y_0)$ at any $y_0$ on the membrane should be a physical velocity. This means $\gamma^2(y_0) > 0$ at all $y_0$ on the membrane.\\

Since we are solving the equation $K^2 = \frac{\gamma^2}{\beta^2}$, it may give solutions with $K^2 <0$ and $\gamma^2 < 0$, which we have to watch out for. If a solution has $K^2 < 0$ then it also has $\gamma^2 <0$, so it would suffice to check the sign of $K^{-2}$.

\subsection{Rotating membranes in flat background}

\hspace{1cm} For the flat space membranes,
$$K^{-2} = D^{-2}(2g + (g'(r))^2),$$
which is positive definite, so it seems like any solution with $2g > 0$ will be acceptable. But if we look at 
$$\gamma^{-2} = 1 - \omega^2 r^2,$$
the non-compact solutions should be unacceptable as $\gamma^{-2}$ goes negative for arbitrarily large value of $r$ for nonzero $\omega$. However on solving the equation \eqref{shapeflt}, it is evident that this equation has only compact solutions of the form
$$2g = \beta^2 - Cr^2,$$
where $C$ is positive, and given by \eqref{csol}. Since $\gamma^{-2}$ given above is a monotonically decreasing function of $r$, its minimum value, which is at the maximum $r$ on the membrane, has to be positive. Since $r^2_{max} = C\beta^2$, 
$$\gamma^{-2}_{min} = 1 - C\omega^2 \beta^2 = 1 - C\Upsilon^2,$$ 
and $C<1$ from \eqref{csol}, so
$$\gamma^{-2}_{min} > 1 - \Upsilon^2.$$
But from \eqref{Upsrange}, $\Upsilon^2 < \frac{1}4$, and hence $\gamma^{-2}_{min} > 0$ for the solutions \eqref{actsol}.

\subsection{Rotating membranes in AdS background}

\hspace{1cm} For the AdS space membranes,
$$K^{-2} = \frac{1}{D^2\left(1+\frac{2g - rg'(r)}{L^2}\right)^2}\left(2g + (g'(r))^2 + \frac{(2g - rg'(r))^2}{L^2}\right),$$

The solutions we consider have the form
$$2g = \beta^2A\mathcal{L}^2 - Cr^2 = A L^2 - Cr^2.$$

Now consider the solutions \eqref{sol1}, i.e. $A = -1$, $C = 0,1$. Since $2g$ has to be positive somewhere, both these solutions are discarded.\\

Now let's consider the branch
$$A\mathcal{L}^2 = (1+A)^2$$
Which means $A > 0$. For $2g > 0$, $K^{-2} > 0$ for any $A$ and $C$, and the solutions should be acceptable. \\

\subsection{Noncompact rotating membranes in AdS} \label{noncomp}

\hspace{1cm} In \ref{rotsolAdS} we analyzed the compact membrane solutions in AdS, i.e. $C > 0$. However there are valid solutions for $C < 0$, i.e. the noncompact solutions which we discuss below.

$$2g = A L^2 - Cr^2$$
represents a noncompact solution that extends from $r = 0$ to $r = \infty$ for $C < 0$. The expressions for $A_\pm$ and $C_{\pm\pm}$ are given in \eqref{sol2} and \eqref{sol3}. These solutions are valid in the parameter ranges that are complementary to the ones in the table \ref{tab1}; those are given in \ref{tab2}. According to the above subsection, these solutions are valid in their respective parameter range.\\

\begin{table}
	\begin{center}
		\begin{tabular}{|l|r|c|}\hline
			\hline
			Parameter range & allowed $A$ & allowed $C$\\
			\hline
			\hline
			$|\Upsilon| < \frac{1}{\mathcal{L}}$ & $A_+$ & $C_{+-}$ \\
			& $A_-$ & $C_{--}$ \\
			\hline
			$\frac{1}{\mathcal{L}} < |\Upsilon| \leq \frac{1}{2}$ & $A_+$ & $C_{++}$ \\
			& $A_+$ & $C_{+-}$\\
			\hline
		\end{tabular}
		\caption{Allowed branches of solutions for different values of $\U$}\label{tab2}
	\end{center}
\end{table}

Naively it looks like for these extended solutions as we go away from $r = 0$, the speed $v = r\omega$ grows larger and it may become superluminal at some point. But in the coordinate system we have chosen, the speed of light also grows larger as $r$ grows. So the membrane never attains superluminal speed.

\section{Validity regimes of the solution for the rotating $AdS$ solutions}	\label{valsolap}
In this appendix we give a derivation of the conditions presented in \eqref{tab1}.\\
For a real solution $\sqrt{1-4\U_i^2}>0$ which gives us the condition 
\begin{equation}
-\frac{1}{2}<\U_i<\frac{1}{2}.
\end{equation}
The compactness and reality condition on $A_{i\pm}$ gives us the condition that 
\begin{equation}\begin{split}
&\CL^2-4>0,\\
&\CL>2\qquad \text{or} \qquad\CL<-2.
\end{split}
\end{equation}
Since $\CL$ is a positive quantity, $\CL>2$.
\begin{equation}\label{amxmin}
A_{+,min}=1 \qquad A_{-,max}=1.
\end{equation}
Hence 
\begin{equation}
A_{i+}-1>0,\qquad A_{i-}-1<0.
\end{equation}
\begin{itemize}
	\item  For $C_{i++}>0$ 
	\begin{equation}\begin{split}
	C_{++}&=\frac{1}{2}\left((1-A_{+}+\CL\sqrt{A_{+}(1-4\U^2)})\right)>0,\\
	&\Rightarrow\CL \sqrt{A_{i+}(1-4\U_i^2)}>(A_{i+}-1),\\
	&\Rightarrow A_{+}\left(\CL+\frac{1}{\U_i}\right)\left(\CL-\frac{1}{\U_i}\right)<0,\\
	&\Rightarrow|\U|<\frac{1}{\CL}.
	\end{split}
	\end{equation}
	\item $C_{i+-}<0$ since both the terms are negative.
	\item $C_{i-+}>0$ since both the terms are positive definite.
	\item For $C_{i--}>0$ 
	\begin{equation}\begin{split}
	C_{--}&=\frac{1}{2}\left((1-A_{-}-\CL\sqrt{A_{+}(1-4\U^2)})\right)>0,\\
	&\Rightarrow \CL \sqrt{A_{i-}(1-4\U_i^2)}<(1-A_{i-}),\\
	&\Rightarrow\left(\CL+\frac{1}{\U_i}\right)\left(\CL-\frac{1}{\U_i}\right)>0,\\
	&\Rightarrow|\U|>\frac{1}{\CL}.
	\end{split}
	\end{equation}
	\section{Coefficients of the perturbative expansion  for the charged  $AdS$ solutions}\label{coeffads}
	The coefficients for the equation \eqref{eqg2}
	\begin{equation}\begin{split}
	a_2&=\frac{2 (A+1) \left(C+\CL^2 \U ^2-1\right)^2}{\CL^2},\\
	b_2&=-2 \left(2 (A+1)^2-A \CL^2\right) \left(C+\CL^2 \U ^2-1\right),\\
	c_2&=-2 A (A+1) \CL^4+2 (A+1)^3 \CL^2-2 C,\\
	a_3&=-\frac{4 (A+1) \left(C+\CL^2 \U ^2-1\right)^2}{\CL^2},\\
	b_3&=2 \left(5 (A+1)^2-(2 A+1) \CL^2\right) \left(C+\CL^2 \U ^2-1\right),\\
	c_3&=2 (A+1) (2 A+1) \CL^4-6 (A+1)^3 \CL^2,\\
	a_4&=\frac{\sigma_n}{\sigma_d},\\
	\sigma_n&=4 (A+1)^4 \left(C+\CL^2 \U ^2-1\right)^2 (3 A^4+12 A^3+2 A^2 \left(-2 \CL^4 \U ^2+\CL^2+9\right)\\&+4 A \left(-2 \CL^4 \U ^2+\CL^2+3\right)-\CL^4 \left((1-2 C)^2+4 \U ^2\right)+2 \CL^2+3),\\
	\sigma_d&=\left((2 A+1) (2 C-1) \CL^4+2 (A+1)^2 \CL^2 (A-3 C+2)-3 (A+1)^4\right)^2,\\
	b_4&= \frac{\xi_n}{\xi_d},\\
	\xi_n&=4 (A+1)^3 \CL^2 \left(C+\CL^2 \U ^2-1\right) (-(A+1) (1-2 C)^2 \CL^6+2 (A+1)^3 (2 C-1) \CL^4\\&+4 (A+1)^4 \CL^2 \left(C+\CL^2 \U ^2-1\right)-4 (A+1)^3 \CL^2 \left((A+1)^2+(2 C-1) \CL^2\right)\\&+2 \left((A+1) (2 C-1) \CL^2+(A+1)^3\right)^2-2 \left((A+1)^2+(2 C-1) \CL^2\right) \bigg((2 (A+1)^2\\&-(2 A+1) \CL^2) ((A+1)^2+(2 C-1) \CL^2)+2 (A+1)^4\bigg)+3 (A+1)^5 \CL^2),\\
	\xi_d&=\left((2 A+1) \CL^2-3 (A+1)^2\right)^2 \left((A+1)^2+(2 C-1) \CL^2\right)^2,\\
	c_4&=\frac{-(A+1)^2 (2 A+1)^2 \CL^8+2 (A+1)^4 \CL^6+3 (A+1)^6 \CL^4}{\left((2 A+1) \CL^2-3 (A+1)^2\right)^2}.
	\end{split}
	\end{equation}
	The coefficients for \eqref{solg2}
	\begin{equation}\begin{split}
	&\zeta_1=\CL^5 (\CL^9-\left(\alpha ^4-10 \alpha ^2+4\right) \CL^7+\left(-10 \alpha ^6+40 \alpha ^4-24 \alpha ^2+2\right) \CL^5\\&+4 \alpha ^2 \left(-8 \alpha ^6+20 \alpha ^4-12 \alpha ^2+1\right) \CL^3-8 \alpha ^8 \left(\alpha ^2-1\right) \sqrt{4 \alpha ^2+\CL^2-4}\\&-2 \alpha ^4 \left(9 \alpha ^4-16 \alpha ^2+5\right) \CL^2 \sqrt{4 \alpha ^2+\CL^2-4}+\CL^8 \sqrt{4 \alpha ^2+\CL^2-4}\\&-\left(\alpha ^4-8 \alpha ^2+2\right) \CL^6 \sqrt{4 \alpha ^2+\CL^2-4}-8 \alpha ^2 \left(\alpha ^4-3 \alpha ^2+1\right) \CL^4 \sqrt{4 \alpha ^2+\CL^2-4}\\&-32 \alpha ^6 \left(\alpha ^2-1\right)^2 \CL)\times\frac{1}{\chi_1},\\
	&\chi_1 =2 \left(-4 \alpha ^6+4 \alpha ^4+\CL^4-\left(\alpha ^2-4\right) \alpha ^2 \CL^2\right) (8 \alpha ^6 \left(\alpha ^2-1\right)+\CL^8+\left(10 \alpha ^2-3\right) \CL^6\\&+3 \alpha ^2 \left(11 \alpha ^2-6\right) \CL^4+\left(38 \alpha ^6-28 \alpha ^4\right) \CL^2+4 \alpha ^4 \left(3 \alpha ^2-1\right) \CL \sqrt{4 \alpha ^2+\CL^2-4}\\&+\CL^7 \sqrt{4 \alpha ^2+\CL^2-4}+\left(8 \alpha ^2-1\right) \CL^5 \sqrt{4 \alpha ^2+\CL^2-4}+\alpha ^2 \left(19 \alpha ^2-4\right) \CL^3 \sqrt{4 \alpha ^2+\CL^2-4})\\
	&\zeta_2=-\frac{\alpha ^4 \CL^3 \left(\CL^3-\CL^2 \sqrt{4 \alpha ^2+\CL^2-4}-2 \alpha ^2 \sqrt{4 \alpha ^2+\CL^2-4}+4 \left(\alpha ^2+1\right) \CL\right)}{\chi_2},\\
	&\chi_2=2 \left(-4 \alpha ^6+4 \alpha ^4+3 \CL^4+\left(-\alpha ^4+12 \alpha ^2+4\right) \CL^2\right) (8 \alpha ^6 \left(\alpha ^2-1\right)+\CL^8+\left(10 \alpha ^2-3\right) \CL^6\\&+3 \alpha ^2 \left(11 \alpha ^2-6\right) \CL^4+\left(38 \alpha ^6-28 \alpha ^4\right) \CL^2+4 \alpha ^4 \left(3 \alpha ^2-1\right) \CL \sqrt{4 \alpha ^2+\CL^2-4}\\&+\CL^7 \sqrt{4 \alpha ^2+\CL^2-4}+\left(8 \alpha ^2-1\right) \CL^5 \sqrt{4 \alpha ^2+\CL^2-4}+\alpha ^2 \left(19 \alpha ^2-4\right) \CL^3 \sqrt{4 \alpha ^2+\CL^2-4}).\end{split}
	\end{equation}
\end{itemize}

\newpage

\bibliographystyle{JHEP}
\providecommand{\href}[2]{#2}\begingroup\raggedright\endgroup

\end{document}